\documentclass[12pt,preprint]{aastex}
\usepackage{psfig}

\begin{document}

\title{Disk-Bulge-Halo Models for the Andromeda Galaxy}

\author{Lawrence M. Widrow, Kathryn M. Perrett, and Sherry H. Suyu}
\affil{Department of Physics, Queen's University, 
Kingston, Ontario, Canada K7L 3N6}

\altaffiltext{1}{widrow@astro.queensu.ca}

\altaffiltext{2}{perrett@astro.queensu.ca}

\begin{abstract}

We present a suite of semi-analytic disk-bulge-halo models for the
Andromeda galaxy (M31) which satisfy three fundamental conditions: (1)
internal self-consistency; (2) consistency with observational data;
and (3) stability of the disk against the formation of a central bar.
The models are chosen from a set first constructed by Kuijken and
Dubinski.  We develop an algorithm to search the parameter space for
this set in order to best match observations of the M31 rotation
curve, inner velocity dispersion profile, and surface brightness
profile.  Models are obtained for a large range of bulge and disk
masses; we find that the disk mass must be $\la 8\times
10^{10}\,M_\odot$ and that the preferred value for the bulge mass is
$2.5\times 10^{10}\, M_\odot$.  N-body simulations are carried out to
test the stability of our models against the formation of a bar within
the disk.  We also calculate the baryon fraction and halo
concentration parameter for a subset of our models and show that the
results are consistent with the predictions from cosmological theories
of structure formation.  In addition, we describe how gravitational
microlensing surveys and dynamical studies of globular clusters and
satellites can further constrain the models.

\end{abstract}


\section{Introduction}

Owing to its size, proximity (distance $\sim 770$ kpc), and long
history of observation, the Andromeda galaxy (M31) offers a unique
opportunity to study in detail the components of a large spiral (Sb)
galaxy.  The aim of this paper is to present new disk-bulge-halo
models for M31 that are (1) internally self-consistent; (2) consistent
with published observations within $30\,{\rm kpc}$; and (3) stable
against the rapid growth of bar-like modes in the disk.  Our models are
drawn from a general class of three-component models constructed by
\citet[][hereafter KD]{kui95}, chosen to fit available observations of
the M31 rotation curve, surface brightness, and bulge velocity
profiles.  The models span a parameter space defined by the disk and
bulge masses, flattening parameter and tidal radius of the galactic
halo.

One advantage of the KD models is that they provide the full
phase-space distribution function (DF) for each of the components.
The DFs are simple functions of three integrals of motion: the energy,
the angular momentum, and an approximate integral which describes the
vertical motions of particles in the disk.  This feature insures that
the models are (very nearly) in dynamical equilibrium.  The KD models
are each specified by 15 parameters.  Of these, 10 have a direct
effect on the rotation curve, velocity dispersion, and surface
brightness profile of the model galaxy.  A search of this parameter
space yields a set of models that fit the observational data to within
quoted uncertainties.

The data considered in this paper constrain the inner $\sim 30\,{\rm
kpc}$ of M31 but say little about the mass distribution at larger
radii.  For this, one must turn to dynamical tracers such as dwarf
satellites \citep{mat98, eva02} and globular clusters \citep{per02}.
Recently \citet{ew00} derived an estimate for the total mass of M31
based on observations of satellites and outer halo globular clusters.
Their analysis assumed simple analytic forms for the gravitational
potential of M31 and the DFs of the tracer populations.  The
models and methods presented in this paper may provide the basis for
future investigations along similar lines and, in particular, a
unified and self-consistent treatment of the full data set associated
with M31.

Two gravitational microlensing surveys toward M31 are currently
underway \citep{cro01, cro96, ker01, cal02}.  Similar surveys toward
the Magellanic Clouds have so far yielded inconclusive results.
Roughly 20 microlensing events have been observed toward the LMC and
SMC \citep{alc00, las00}, but one cannot say whether the lenses
responsible for these events are indeed MACHOs or simply stars within
the Magellanic Clouds or Milky Way disk.  The M31 microlensing
experiments should be able to resolve this question.  M31 is highly
inclined and therefore lines of sight toward the far side of its disk
pass through more of its halo than do lines of sight toward the near
side.  If the halo of M31 is composed (at least in part) of MACHOs,
there will be more microlensing events occurring toward the far side
of the galaxy \citep{cro92}.  Previous efforts to compute theoretical
optical depth and event rate maps for M31 assumed ad hoc models for
the disk, bulge, and halo.  Our models represent a significant
improvement over these models since they are both internally
self-consistent and are consistent with published data on M31.  We
will describe how one can compute optical depth and event rate maps
for the models constructed in this study.

This paper takes the following form.  In Section 2 we review published
observations of the rotation curve, velocity dispersion profile, and
surface brightness profile of M31.  Section 3 provides a summary of
the main features of the KD models.  In Section 4, we describe a
method to search parameter space for models which best fit the
observations.  The results of this search are presented in Section 5.
Promising models are found which span a wide range in disk mass and in
halo tidal radius and shape.  We compute various properties of these
models such as the disk and bulge mass-to-light ratios and the baryon
mass fraction.  For a subset of the models, we calculate the mass
distribution and line-of-sight velocity dispersion profiles.  We also
fit the density profile to the NFW profile \citep{nfw96} and thus
obtain an estimate for the halo concentration parameter.  In Section
6, we describe a technique to construct theoretical event rate maps
and present one example.  Our conclusions and a discussion of possible
extensions of this work are presented in Section 7.

\section{Observations}
\label{sec:OBS}

Our analysis utilizes published measurements of the galaxy's rotation
curve, average surface brightness profile, and bulge velocity
profiles.  There is a considerable amount of M31 observational data of
dissimilar quality available from the literature.  In this section, we
briefly describe the data selected for use in fitting the KD models.

\subsection{Rotation Curve}

Rotation curves for the M31 disk have been obtained from optical and
radio observations spanning various ranges in galactocentric radius
\citep[e.g., ][]{rub70, rub71, got70, deh75, ken89, bra91}.  An
optimal combination of such data sets requires a good understanding of
any associated calibration errors and uncertainties.  In this study,
the composite rotation curve for the galaxy was obtained by combining
velocity data from the studies of \citet{ken89} and \citet{bra91}.
\citet{ken89} obtained velocities with estimated statistical errors of
$\sim 6\,{\rm km\,s^{-1}}$ for 30 HII emission regions along the major
axis of the galaxy, with galactocentric radii in the range of
6\,--\,25\,kpc.  The \citet{bra91} measurements of neutral hydrogen
within the gaseous disk of M31 yielded velocity measurements out to a
radius of $r\sim30$~kpc.  Braun's data within 2~kpc of the M31 center
were neglected here due to possible distortions arising from the
presence of a bar-like triaxial ellipsoidal bulge.  Beyond
$r\sim20$~kpc, measurements were obtained for spiral arm segments on
only one side of the galaxy, hence data in this region were also not
incorporated into our fitting.  Details of the rotation curve
interpolation can be found in \citep{bra91}.

\begin{figure}     
\plotone{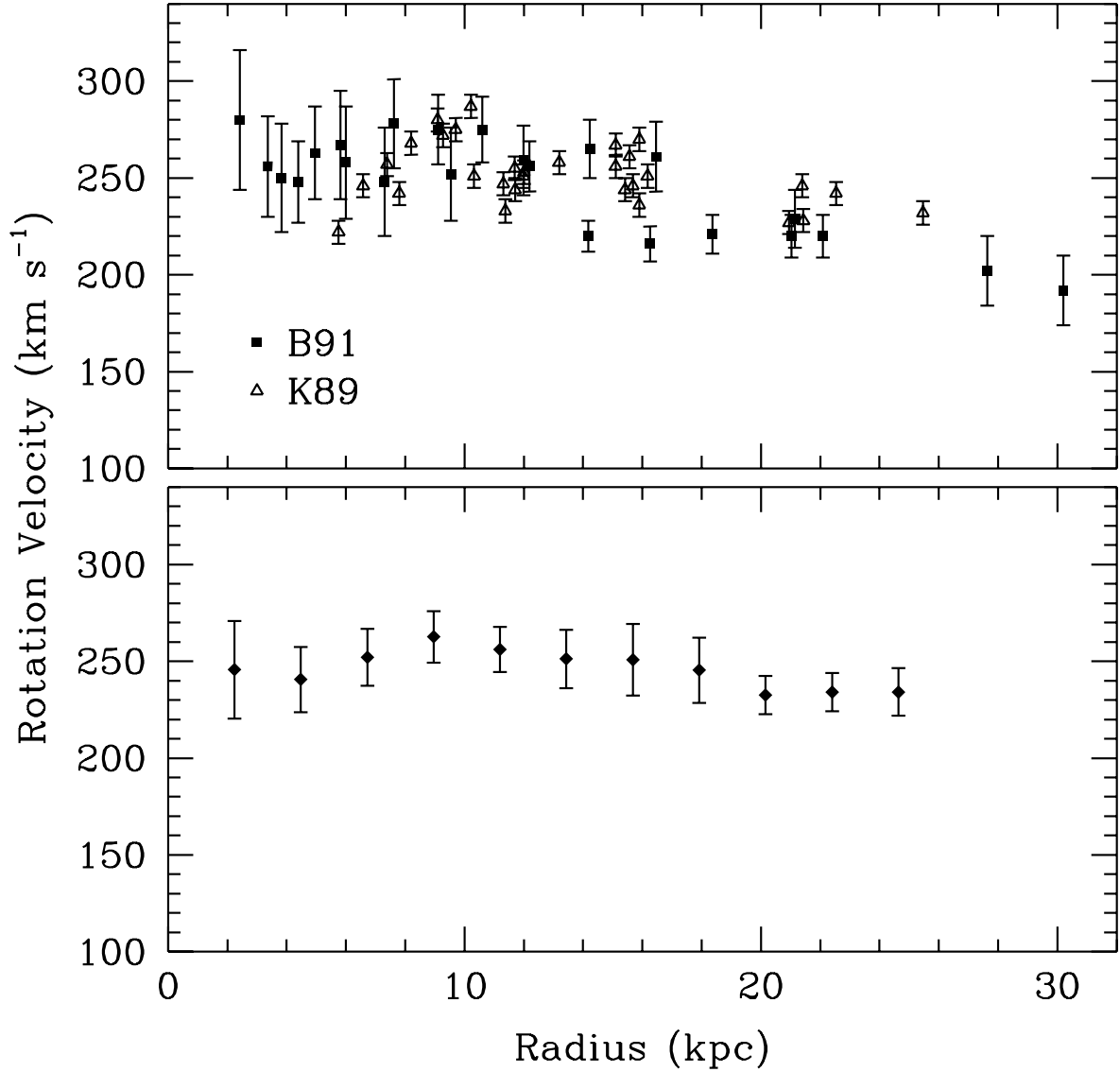}
\caption{The disk rotation curve.  The rotation measurements of
\citet[][``K89'']{ken89} and \citet[][``B91'']{bra91} are shown in the
upper panel, and the smoothed rotation profile used in the model
fitting is provided in the lower panel.}
\label{fig:rotation}  
\end{figure}

Both sets of rotation velocity measurements and their respective
errorbars are shown in the upper panel of Figure~\ref{fig:rotation}.
Kernel smoothing was used to form the composite disk rotation curve,
which is shown with its RMS uncertainties in the lower panel of
Figure~\ref{fig:rotation}.

In determining the rotation curve of the galaxy, Braun assumes
circular gas motions within the disk.  It should be noted that the
presence of certain dynamical anomalies may lead to systematic errors
in the rotation curve determination.  Such anomalies will be discussed
at the end of Section~\ref{sec:DBHModels}.
  

\subsection{Surface Brightness Profile}

Although there have been many optical photometric studies of M31
\citep[see, for example, Table~4.1 of][]{wal87}, the task of combining
such data sets is not straightforward.  Differences in filter
bandpasses and resolution between the various studies yield systematic
errors which are generally difficult to characterize.  Discrepancies
in the adopted galaxy inclination and isophote orientation also
contribute to differences in the light profiles obtained by different
authors.  For these reasons, we opted to avoid combining different
data sets and instead adopted the surface brightness profile from
\citet{wal87}.  These authors produced a global light profile for M31
out to $r\sim28$~kpc by averaging the distribution of galaxy light
over elliptical rings, assuming an inclination of $77^\circ$.  The
inner parts of the galaxy are dominated by light from the bulge
component, which itself has a significantly higher position angle (PA)
than that of the disk: ${\rm PA}\sim 50^\circ$ versus $38^\circ$
\citep{hod82}.

There is a distinct variation in the position angle of elliptical
isophotes fitted to the surface brightness of M31 as one looks out in
galactocentric radius \citep{hod82}.  Furthermore, there is a
significant warp in the disk beyond $r\sim 22\,{\rm kpc}$
\citep{wal87}.  These structural features cannot be reproduced in the
KD models and will thus contribute to fitting errors calculated for
the surface brightness profiles.

\subsection{Bulge Velocity Profiles}

\begin{figure}     
\plotone{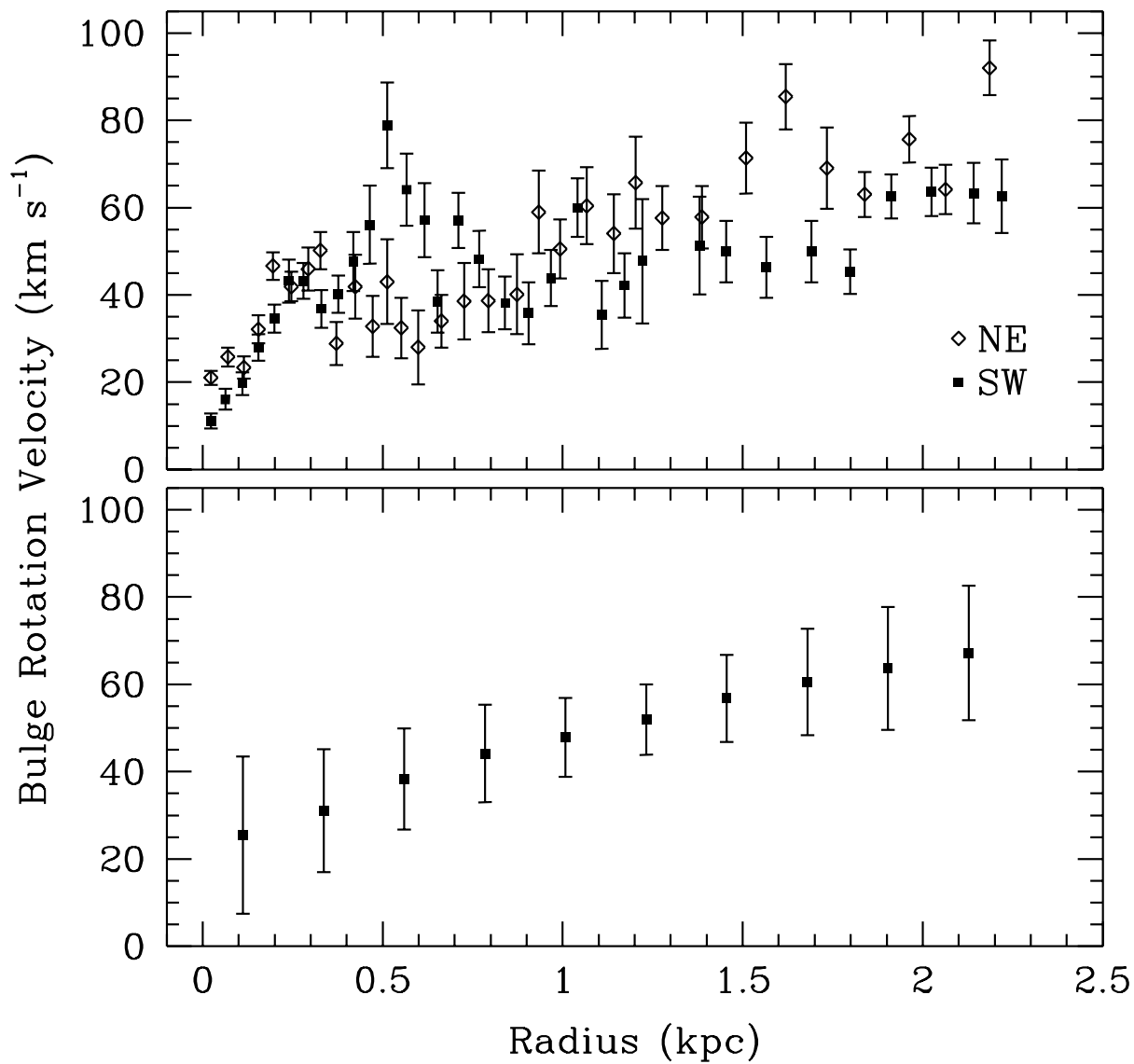}
\caption{Bulge rotation velocity as a function of radius.  The upper
panel shows the bulge rotation measurements of \citet{mce83} for the
north-east (NE) and southwest (SW) sides of the minor axis along an
axis with a position angle of $45^\circ$.  The smoothed bulge rotation
curve data and errors used in the model fitting are shown in the lower
panel.}
\label{fig:brotation}  
\end{figure}

\begin{figure}     
\plotone{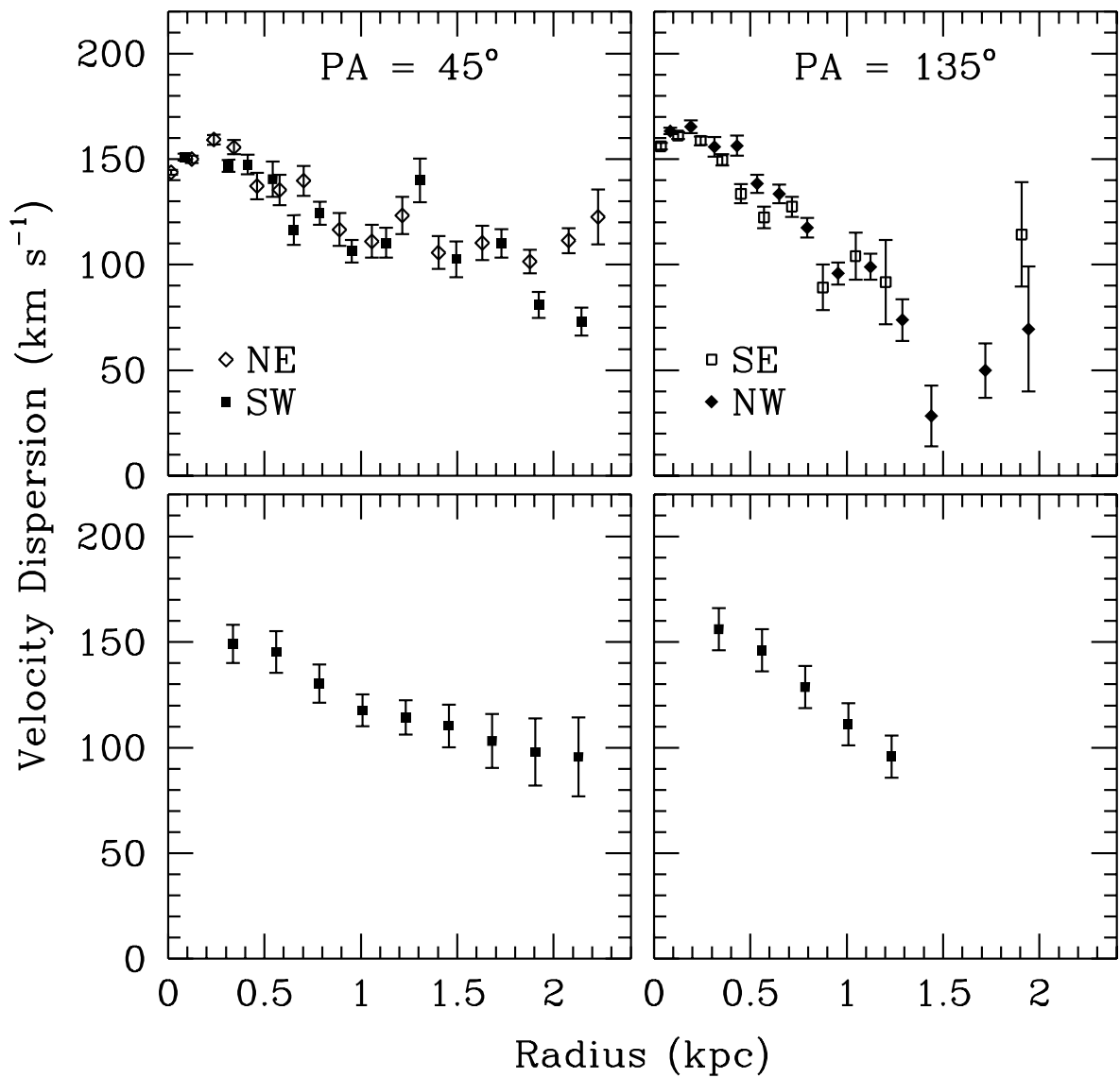}
\caption{Velocity dispersion measurements of \citet{mce83} along the
major axis (upper-left panel) and minor axis (upper-right panel) of
the bulge.  The corresponding smoothed profiles used in the model
fitting are shown in the lower panels.  The four outer points in the
minor axis data were neglected due to their large uncertainties.} %
\label{fig:dispersion}  
\end{figure}

The dynamics of the bulge can be used to deduce the mass distribution
in that component of the galaxy.  We utilize the stellar rotation and
velocity dispersion results of \citet{mce83} along the bulge major
axis (${\rm PA}=45^\circ$) and minor axis (${\rm PA}=135^\circ$) in
the fits to the KD models described later in
Section~\ref{sec:DBHModels}.  The bulge data were smoothed using the
same kernel averaging technique mentioned above; the rotation and
dispersion results are shown in Figures~\ref{fig:brotation} and
\ref{fig:dispersion}, respectively.

\citet{mce83} noted several asymmetries in the rotation curves along
various position angles of the bulge for $|R| < 10^\prime$, although
the cause of these asymmetries was unclear.  The KD models are unable
to reproduce such asymmetries within the bulge; we will return to this
point later in the next section.

\section{Self-Consistent Disk-Bulge-Halo Models}
\label{sec:DBHModels}

\citet{kui95} constructed a set of semi-analytic models for the
phase-space DFs of disk galaxies.  Their models have three components:
a thin disk, a centrally concentrated bulge, and an extended halo.  In
this section we summarize the essential features of these models.

The halo DF is taken to be a lowered Evans model \citep{kui94}.  Evans
models \citep{eva93} are exact, two-integral distribution functions
for the flattened logarithmic potential, $\Psi\propto \log{\left
(x^2+y^2+z^2/q^2\right )}$ where $q$ is the flattening parameter.  As
with the isothermal sphere, Evans models are infinite in extent.  In
analogy with the lowered isothermal sphere or King model
\citep{kin66}, lowered Evans models introduce a truncation in energy
such that the system that results has finite mass.  The halo DF

\begin{equation}
\label{eq:haloDF}
f_{\rm halo}\left (E,\,L_z\right ) = \left\{
	\begin{array}{ll}
		\left [\left (AL_z^2+B\right )
		\exp{\left (-E/\sigma_0^2\right )}+C\right ]
		\left (\exp\left (-E/\sigma_0^2\right )-1\right )
		 & \mbox{if $E<0$} \\
		0 & \mbox{otherwise}\end{array} \right. 
\end{equation}
has five free parameters: $A,\,B,\,C,\,\sigma_0,\,$ and $\Psi_0$, the
central potential for the system.  Following KD, the first three
parameters are replaced by $q$, a characteristic halo radius $R_a$,
and a core radius $R_c$.  In practice, the latter is described in
terms of a core smoothing parameter $\left(R_c/R_K\right )^2$, which
specifies the ratio of the core radius to the derived King radius for
the halo.  The DF is independent of the sign of $L_z$ and may
therefore be written as the sum of two components, one with positive
$L_z$ and the other with negative $L_z$.  Rotation can be incorporated
into the model by varying the relative ``weight'' of these two parts
as specified through the streaming fraction $S_h~(= 1/2~{\rm
for~no~rotation})$.

The bulge DF is given by the lowered isothermal sphere or King model
\citep{kin66, bin87}, and takes the form

\begin{equation}
\label{eq:bulgeDF}
 f_{\rm bulge}\left (E\right ) = \left\{
	\begin{array}{ll}
		\rho_b\left (2\pi\sigma_b^2\right )^{-3/2}
		\exp{\left [\left (\Psi_0-\Psi_c\right )/\sigma_b^2\right ]}
		\left (
		-\exp{\left [\left (E-\Psi_c\right )/\sigma_b^2\right ]}
		-1\right ) & \mbox{if $E<\Psi_c$} \\
		0 & \mbox{otherwise}\end{array} \right. 
\end{equation}
where $\rho_b$ and $\sigma_b$ govern the central density and velocity
dispersion of the bulge.  The cut-off potential, $\Psi_c$, controls
the extent of the bulge.  As with the halo, there is an additional
parameter, the bulge streaming fraction $S_b$, which controls the
rotation of the bulge.

The disk DF depends on three integrals of motion: $E$, $L_z$, and an
approximate third integral corresponding to the energy associated with
vertical oscillations of stars in the disk.  The disk DF can be chosen
to yield a density field with the desired characteristics.  In the KD
models, the density falls off approximately as an exponential in the
radial direction and as ${\rm sech}^2$ in the vertical direction.
Five parameters are used to characterize the disk: its mass $M_d$, the
radial scale length $R_d$, the scale height $h_d$, the disk truncation
radius $R_o$, and the parameter $\delta R_o$ which governs the
sharpness of the truncation.

In a given potential, the DF for each component implies a unique
density field.  For self-consistency, the density-potential pair must
satisfy the Poisson equation.  This is accomplished by an iterative
scheme as described by \citet{kui95}.  The essential point is that the
properties of the density fields of the bulge and halo are implicit
rather than explicit functions of the input parameters.  In
particular, the masses of the bulge and halo ($M_b$ and $M_h$,
respectively) and the halo tidal radius ($R_t$) are determined {\it a
posteriori}.  Likewise, the shapes of the bulge and halo are
determined via the Poisson-solving algorithm.  In particular, the
bulge is flattened due to the contribution to the potential from the
disk.

By design, the KD models are axisymmetric and therein lies the main
limitation to the development of a truly realistic model for M31.
Observational studies indicate that M31 cannot be described adequately
by an axisymmetric model.  Dynamical anomalies for M31 include factors
such as the presence of a triaxial bulge \citep{sta77, ger86},
perturbations caused by its companions M32 and NGC 205 \citep{byr78,
sat86}, the effects of significant variations in the inclination of
the galactic plane as a function of radius \citep{hod82}, areas of
infall motion towards the galaxy center \citep{cra80}, and local
anomalies attributable to fine structure and shocks within the spiral
arms of the galaxy \citep{bra91}.  Small local perturbations in
spectra obtained through dust lanes in the central regions of the
galaxy may also be a factor.  These dust patches would have the effect
of slightly increasing the local radial velocity measurements, thereby
inducing local errors in the rotation curve of the bulge
\citep{mce83}.  Furthermore, the bulge dispersion may be affected by
residual rotation caused by disk contamination along its minor axis.


\section{Fitting KD Models to the Observations}
\label{sec:KD}

The KD models are specified by 15 parameters: four for the bulge, five
for the disk and six for the halo.  Our goal is to determine the
parameter set that yields a model which best fits the observations.
This is accomplished by minimizing the composite $\chi^2$ statistic
that is calculated by comparing the model rotation curve, surface
brightness profile, and bulge velocity profiles with the data sets as
described in Section~\ref{sec:OBS}.

\subsection{Minimization Strategy}

Unfortunately, the data considered in this paper are not sufficient to
determine a global minimum in the full 15-dimensional parameter space.
We therefore adopt a strategy in which a best-fit model is found for
targeted values of the disk and bulge masses, flattening parameter,
and in some cases, tidal radius.  In addition, we do not attempt to
minimize over the disk thickness, disk truncation radius, or halo
streaming fraction (i.e., the parameters $R_o$, $\delta R_o$, $h_d$,
or $S_h$).  Our analysis proceeds as follows:

\begin{enumerate}

\item The surface brightness profile at radii where the disk dominates
is very nearly exponential.  In the R-band, the scale radius of the
disk is $5.4\,{\rm kpc}$.  We therefore fix $R_d$ to this value at the
outset.

\item Since the data used in our analysis have little to say about the
parameters $R_o$, $\delta R_o$ and $h_d$, we set them to be typical
values for M31 of $40\,{\rm kpc}$, $1\,{\rm kpc}$, and $300\,{\rm
pc}$, respectively.  In addition, we assume that the halo does not
rotate, i.e., $S_h=0.5$.

\item We consider models with disk and bulge masses in the following
ranges:
\begin{eqnarray*}
3\times 10^{10}\,M_\odot \le M_d \le 16\times 10^{10}\,M_\odot \\ 
1\times 10^{10}\,M_\odot \le M_b \le 4\times 10^{10}\,M_\odot~.
\end{eqnarray*}
$M_d$ is an input parameter but $M_b$ is a complicated function of
$\rho_b$, $\sigma_b$, $\Psi_c$ and, to a lesser extent, the other
parameters.  Therefore, to force the minimization routine to select a
model with the desired $M_b$, we minimize a ``pseudo-$\chi^2$
statistic'' that includes the term \mbox{$\left ( M_b\left ({\rm
desired}\right ) - M_b\left ({\rm model}\right )\right
)^2/\sigma_{M_b}^2$}.  The user-specified parameter $\sigma_{M_b}$
controls the accuracy with which $M_b$ is fit to the desired value.
Of course, the physically relevant quantity is the $\chi^2$ statistic
associated with the fits to the observational data, and this is the
one quoted herein.

\item As with $M_b$, the tidal radius of the halo is an implicit
function of the other input parameters.  Therefore, for those cases
where we wish to specify the tidal radius, an additional term
\mbox{$\left (R_t\left ({\rm desired}\right ) - R_t\left ({\rm
model}\right )\right )^2/\sigma_{R_t}^2$} is included as part of the
pseudo-$\chi^2$ statistic in order to drive $R_t$ to the desired
value.  Again, the user specifies the value of $\sigma_{R_t}^2$.

\item The flattening parameter $q$ governs the shape of the halo
potential and is specified explicitly.  The shape of the mass
distribution of the halo is determined implicitly through the
Poisson-solving algorithm.

\end{enumerate}
A given model is constructed by minimizing the pseudo-$\chi^2$
statistic for the fit.  This minimization procedure is described in
the next section.

\subsection{Multidimensional Minimization Technique}

Minimization of the pseudo-$\chi^2$ statistic over the
multidimensional KD parameter space is performed by employing the
downhill simplex algorithm \citep[see, for example,][]{pre86}.  An
N-dimensional simplex is a geometrical object consisting of $N+1$
points or vertices and all of the line segments that connect them.
Thus, a simplex encloses a finite volume in an N-dimensional space.
For the case at hand, the space is defined by the 10 free parameters
as described above.  An initial guess at the values of these
parameters fixes one vertex of the initial simplex.  The remaining
vertices of the initial simplex are constructed by stepping in each
direction of parameter space by some appropriate distance, which is
typically set to 10\% of the parameter value.

The downhill simplex algorithm proceeds through a series of iterations
as follows.  The pseudo-$\chi^2$ statistic is calculated at each
vertex of the simplex.  The algorithm then reflects the vertex with
the highest value of pseudo-$\chi^2$ through the opposite face of the
simplex to search for a lower function value.  If a lower value is
found at this new location, the algorithm proceeds by testing the
point twice as far along this line.  The vertex in the original
simplex with the highest pseudo-$\chi^2$ is then replaced by the
reflected position with the lower function value.  If a lower value is
not found at the reflected position, the simplex is contracted about
its vertex with the lowest function value.  In this manner, the
simplex steps through parameter space and gradually contracts around
the point with the minimum pseudo-$\chi^2$, thereby honing in on the
best fit to the observed data and the targeted values for the
component masses and/or tidal radius.

The downhill simplex method has a number of advantages over
minimization procedures that are based on gradients of the function
(e.g., the method of steepest descent; see \citet{pre86} and
references therein).  Gradient methods appear to be more susceptible
to the presence of local minima in the complicated $\chi^2$ surface.
In addition, the simplex method is computationally efficient,
requiring relatively few function evaluations.  Between 10 and 100
iterations are needed in order to locate a position in parameter space
with the minimum value of pseudo-$\chi^2$.  An iteration requires
between 1 and 10 function evaluations, each of which takes a minute or
so of CPU time.  Therefore, a model can typically be generated within
$1-2$ CPU-hours using a standard 1~GHz desktop computer.


\section{The Models}

We begin with a survey of models in the $M_d-M_b$ plane focusing on
the quality of the fits to the observational data, the stability of
the disk, and the mass-to-light ratios of the disk and bulge
components.  We next consider constraints on the mass distribution
beyond $30\,{\rm kpc}$ from dynamical studies of satellite galaxies,
globular clusters, and planetary nebulae.  We also check whether our
preferred models are consistent with predictions from the baryonic
Tully-Fisher relation and cosmological constraints such as the baryon
fraction.  Toward this end we construct a series of models with tidal
radii between $80\,{\rm kpc}$ and $160\,{\rm kpc}$ and also explore
the implications of replacing the lowered Evans halo of a KD galaxy,
which has a sharp cut-off in density at the tidal radius, with an NFW
halo.  Finally, we investigate models with flattened halos, such that
$q<1$.

\subsection{The Disk and Bulge Mass Models}

We have constructed and analysed over twenty M31 models in the
$M_b-M_d$ plane with $q=1$ and $R_t$ unconstrained.
Table~\ref{tab:models} summarizes the results for a select subset of
these models.  In addition to the disk and bulge masses,
Table~\ref{tab:models} provides the composite $\chi^2$ statistic for
the fits to the observational data (column 4), the R-band
mass-to-light ratios for the disk and bulge (columns 5 and 6), the
mass interior to a sphere 30~kpc in radius (column 7), the mass of the
halo (column 8) and the tidal radius (column 9).

\begin{figure}
\plottwo{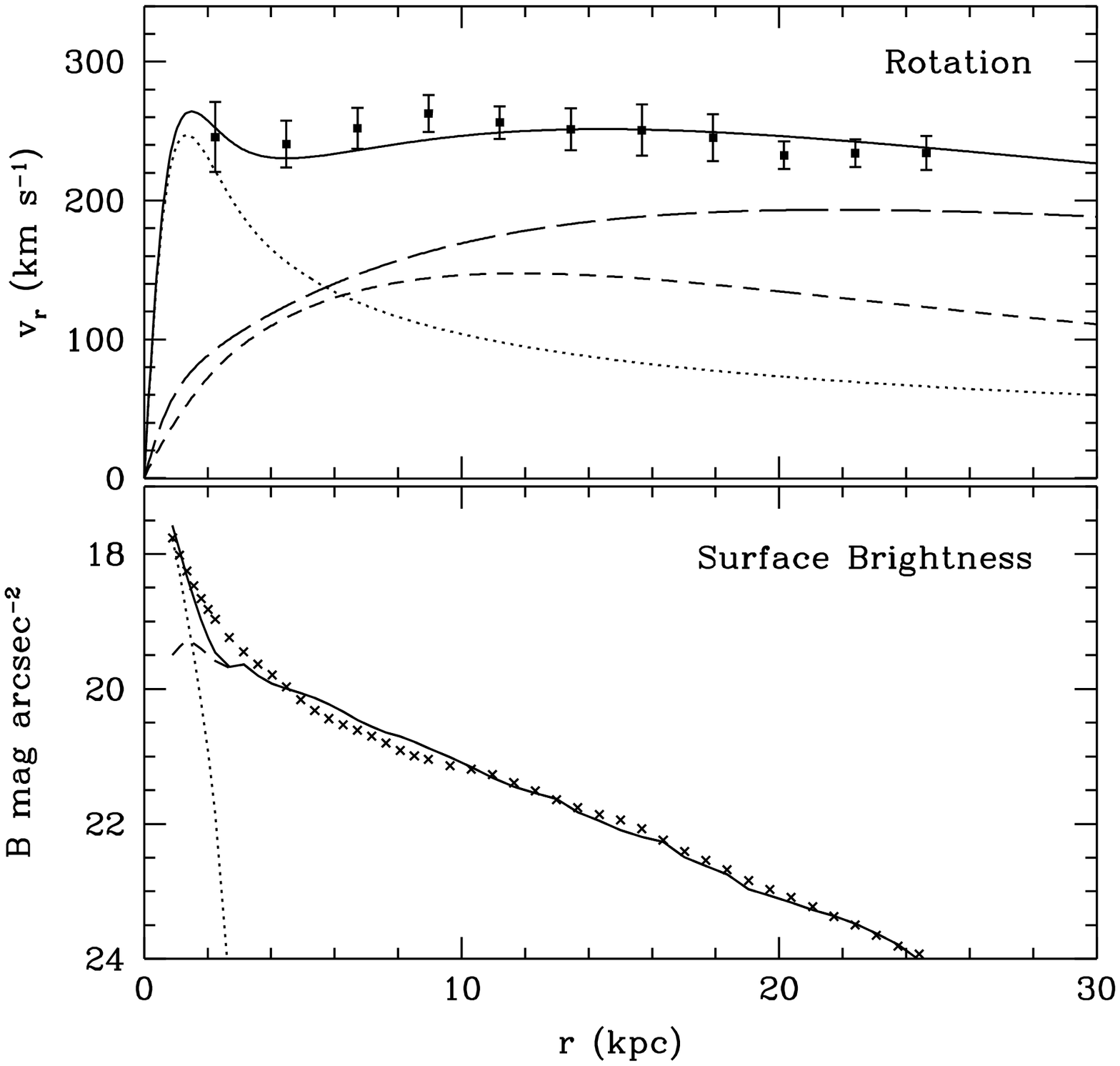}{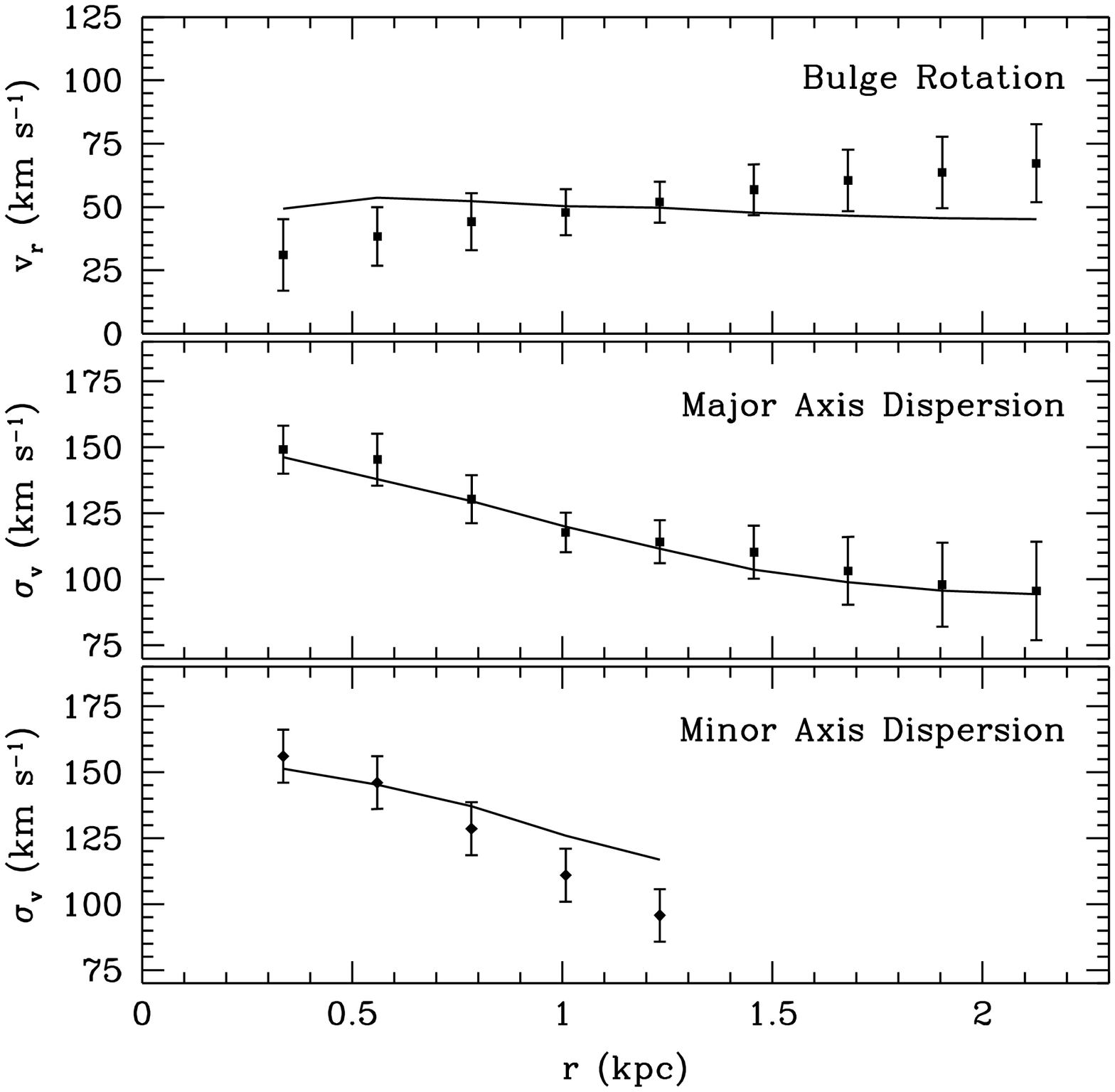} 
\caption{Comparison of theoretical fits with observational data for
Model A from Table~\protect\ref{tab:models}.  The upper left-hand
panel shows the net rotation curve for the galaxy (solid line), along
with the profiles for the individual components: bulge (dotted line),
disk (dashed line) and halo (long-dashed line).  The data points and
error bars are those of the smoothed rotation profile given in
Figure~\protect\ref{fig:rotation}.  The lower left-hand panel shows
the surface brightness profile measurements with the model fits of the
bulge (dotted line), disk (dashed line) and total light (solid line).
The upper-right panel provides the bulge velocity profile data and
model fit.  The middle- and lower-right panels show the inner velocity
dispersion data and resulting fits along the galaxy's major and minor
axes, respectively.}
\label{fig:A}
\end{figure}

\begin{figure}
\plottwo{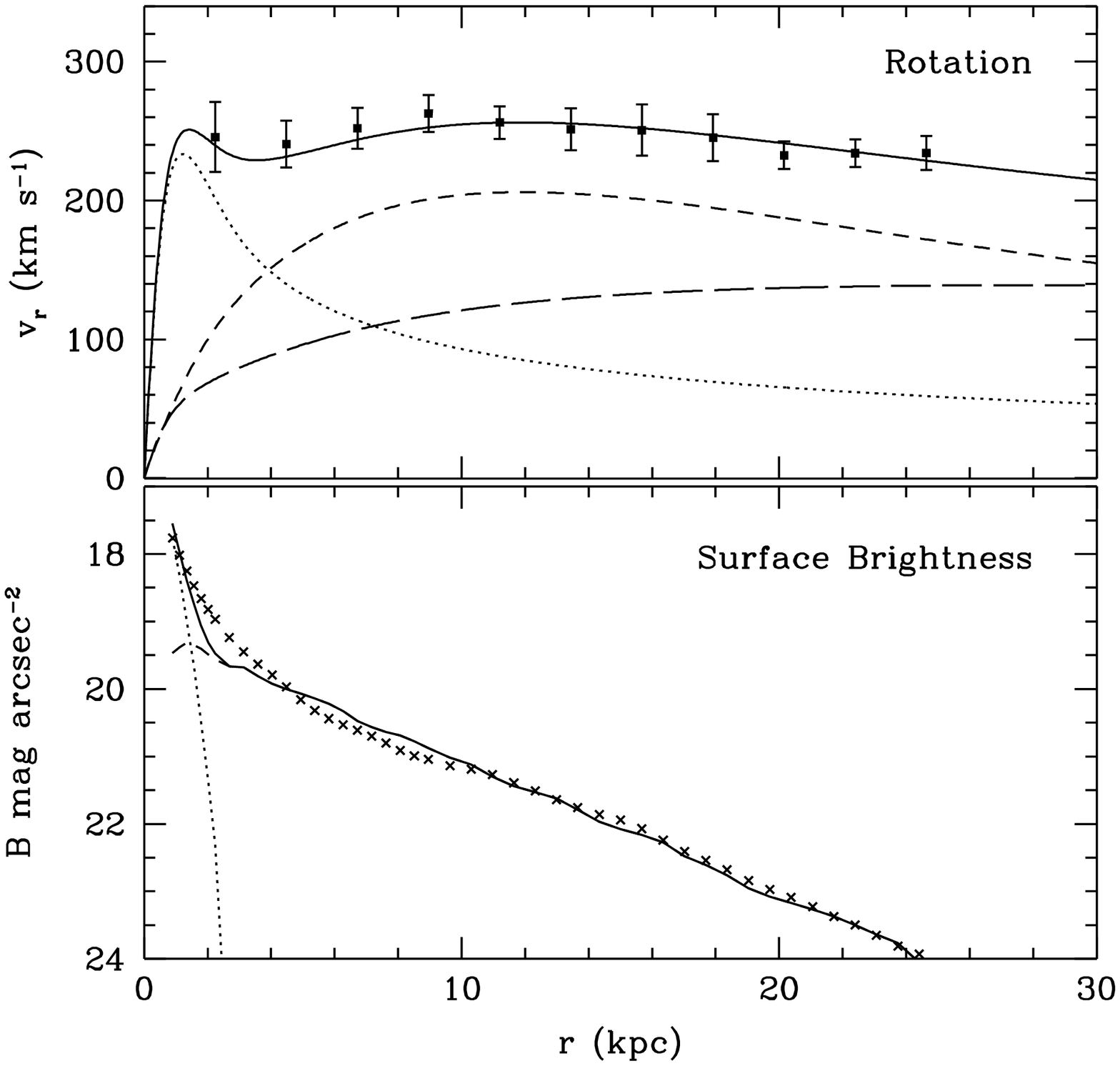}{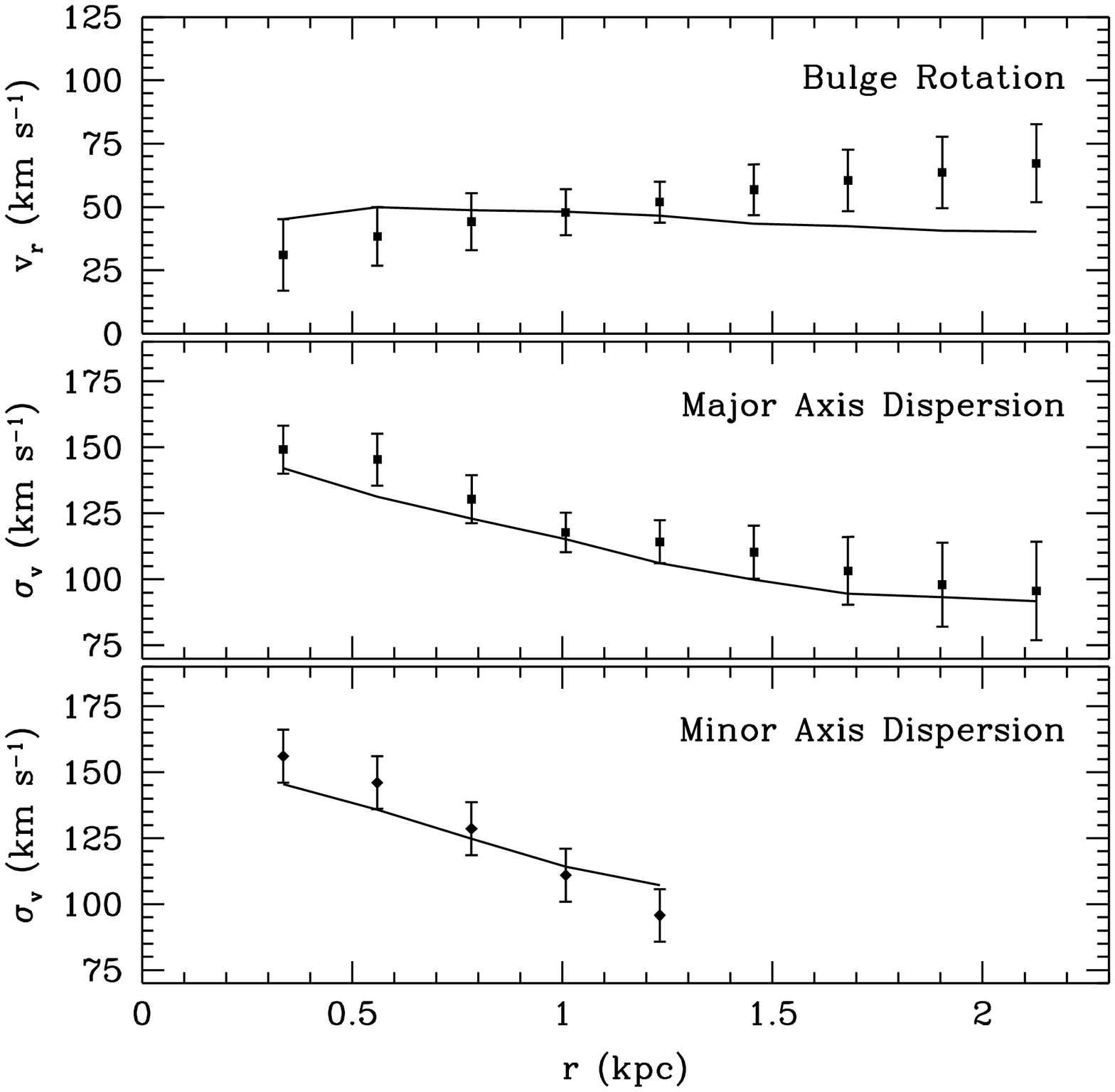}
\caption{Comparison of theoretical fits with observational data for
Model D (see Table~\protect\ref{tab:models}).  The plots and line
types are as described previously in Figure~\protect\ref{fig:A}.}
\label{fig:D}
\end{figure}

Our analysis indicates that there is a trough in $\chi^2$ running
roughly parallel to the \mbox{$M_d$-axis} and centered on $M_b\simeq
2.5\times 10^{10}\,M_\odot$.  Both Models A and D lie near the minimum
of this trough while Models A, B, and C trace out the cross section of
the trough at $M_d=7\times 10^{10}\,M_\odot$.  Models along the
minimum of the trough have low values of $\chi^2$ (typically
$\chi^2\simeq 0.6-1$) indicating an excellent overall fit to the
observations.  This point is illustrated in Figures \ref{fig:A} and
\ref{fig:D} where we compare the theoretical predictions for Models A
and D with the observational data.  The agreement is particularly good
for the gas rotation curve and inner velocity dispersion profiles
along the galaxy's major and minor axes.  Furthermore, the exponential
disk does an excellent job of fitting the surface brightness profile
beyond $5\,{\rm kpc}$ though the fit is not as good in the transition
zone between bulge and disk dominated regions of the galaxy ($2\,{\rm
kpc}\la r\la 4\,{\rm kpc}$).  Moreover, the inner rotation curve for
the model appears to have the wrong shape: the model curves are
relatively flat while the data suggest a rotation curve that is rising
slowly.  These two discrepancies may, in part, reflect the fact that
our models assume an axisymmetric bulge whereas the actual bulge of
M31 is triaxial.  We note that the mass distribution in the bulge
component of the model is flattened --- but still axisymmetric --- due
to its interaction with the gravitational potential of the disk.  The
major-to-minor axis ratio is found to be $\sim 0.8$, in good agreement
with the value found by \citet{ken89}.

Figures \ref{fig:A} and \ref{fig:D} help to explain the existence of
the trough in $\chi^2$.  The contributions to the outer rotation curve
($r\ga 5\,{\rm kpc}$) from the halo and disk are both rather broad in
radius and therefore one can be played off the other.  

Figure \ref{fig:D} shows that the disk dominates the rotation curve of
Model D for $4\,{\rm kpc}\la r\la 30\,{\rm kpc}$.  This
feature, common to all high-$M_d$ models reveals a fundamental problem
with these models, namely a susceptibility to the bar instability.  A
dynamically cold, self-gravitating disk is unstable to bar formation
\citep{hoh71} and since a strong bar instability completely disrupts
the disk, any model in which one is present is unacceptable.

\begin{figure}     
\plotone{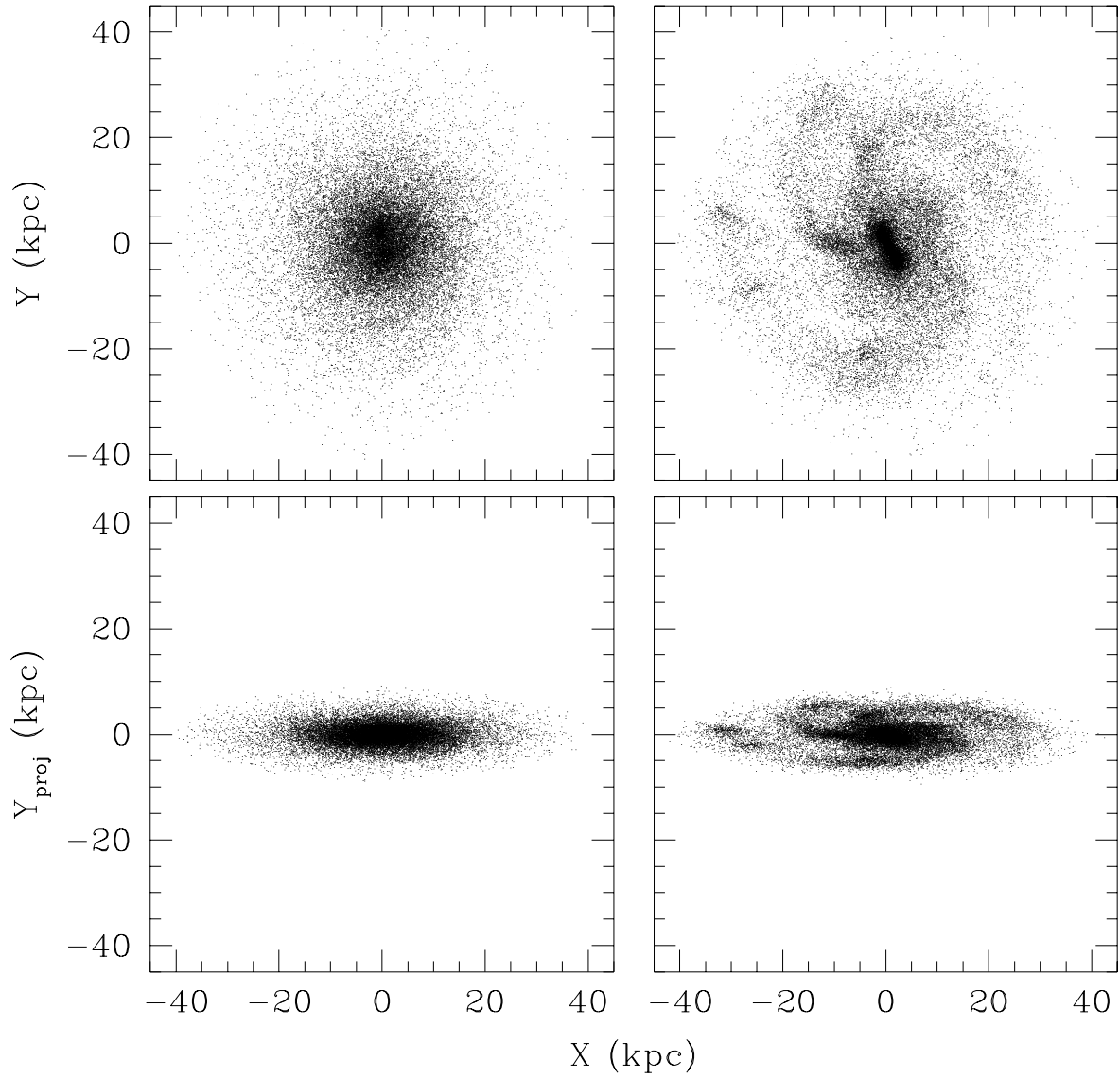}
\caption{Results of N-body simulations for Models A and D after
several dynamical times.  Upper panels provide face-on views of the
disk for Model A (left) and Model D (right).  Lower panels show the
corresponding ``observer's view'' with the disk tilted to an
inclination angle $77^\circ$.}
\label{fig:nbody}  
\end{figure}

In general, bar instabilities can be suppressed by an extended halo, a
bulge that dominates the dynamics of the inner part of the galaxy,
significant vertical velocities among the disk stars, or a combination
thereof \citep{ost73,sel85}.  We have performed a series of N-body
experiments to test the stability of our models.  We use the algorithm
of \citet{deh02}, which has the advantage over tree and mesh codes in
that the computation time scales as the number of simulation
particles, $N$, rather than $N\ln{(N)}$.  The simulations were
performed with $2\times 10^5$ particles for each of the three
components of the model and were run for 8 dynamical times as measured
at one scale radius.  Model D develops a bar within a few dynamical
times as expected given that the disk dominates the gravitational
potential within a radius of $r\sim 30\,{\rm kpc}$ (see Figure
\ref{fig:D}).  Model A, in which the inner rotation curve is dominated
by the bulge and the outer rotation curve is dominated by the halo
(Figure \ref{fig:A}), appears to be stable.  These results are
illustrated in Figure \ref{fig:nbody} where we show face-on and
Milky-Way observer views of the evolved disk-particle distribution for
Models A and D.  Further simulations suggest that the demarcation
between stable and unstable models occurs in the neighborhood of
$M_d\simeq 8\times 10^{10}\,{\rm M}_\odot$.  Models with a disk in
this mass range show signs of weak spiral and bar-like structures
\citep{sta77}.  The presence of spiral structure in the disk of M31 as
well as a triaxial bar-like bulge suggests that weak instabilities
operate in this galaxy.  Thus, $M_d=8\times 10^{10}\,{\rm M}_\odot$
should not be interpreted as a strict upper bound on the disk mass of
M31 but rather as an interesting region of parameter space in which
dynamical evolution may give rise to non-axisymmetric structures
similar to what is observed in M31.  What we can say for sure is that
models like D and K1 are violently unstable and therefore ruled out.
Conversely, Model K2 may be so stable that the mechanisms which
drive spiral structure are suppressed \citep{sel85}.

\begin{figure}
\plottwo{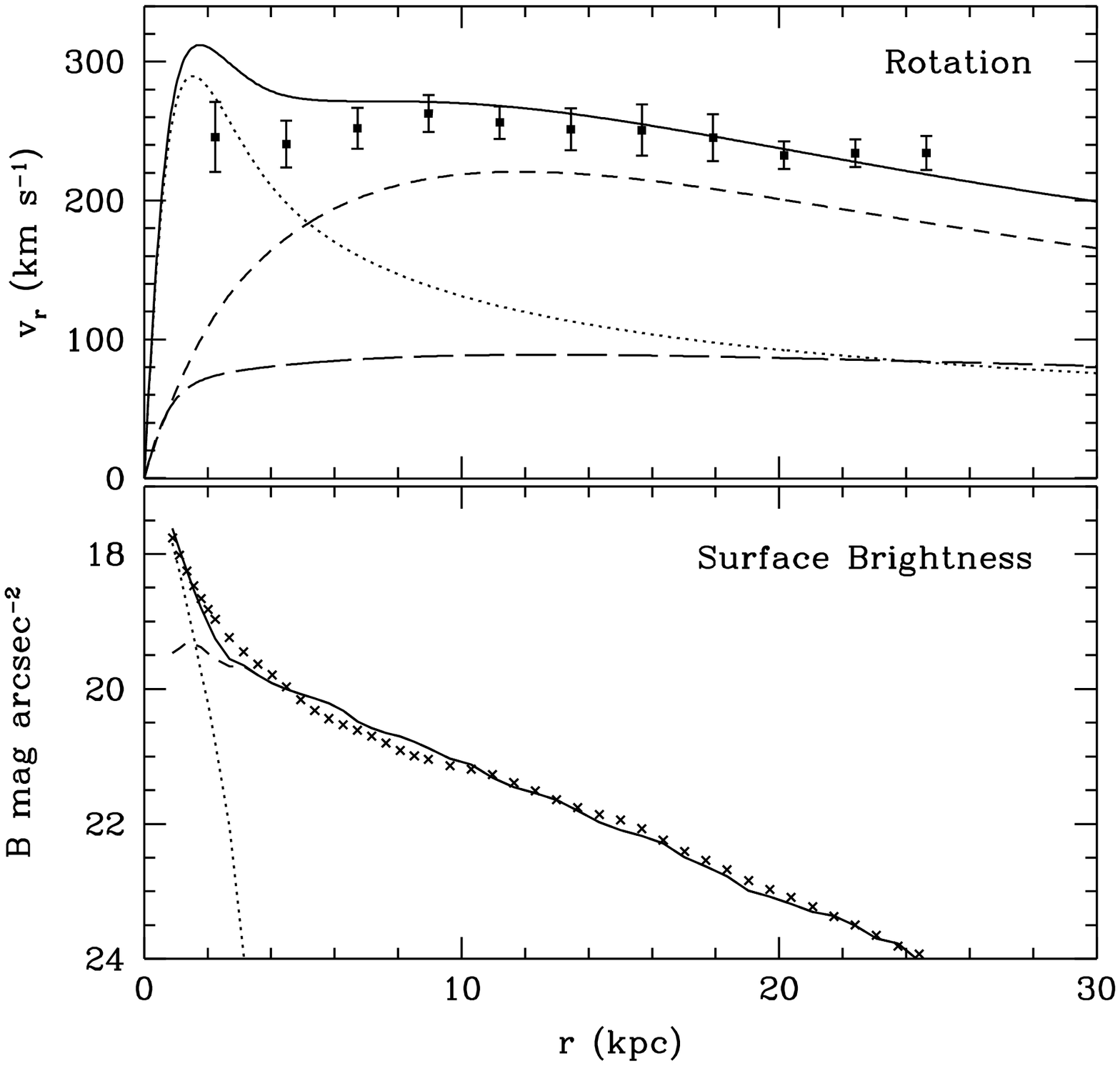}{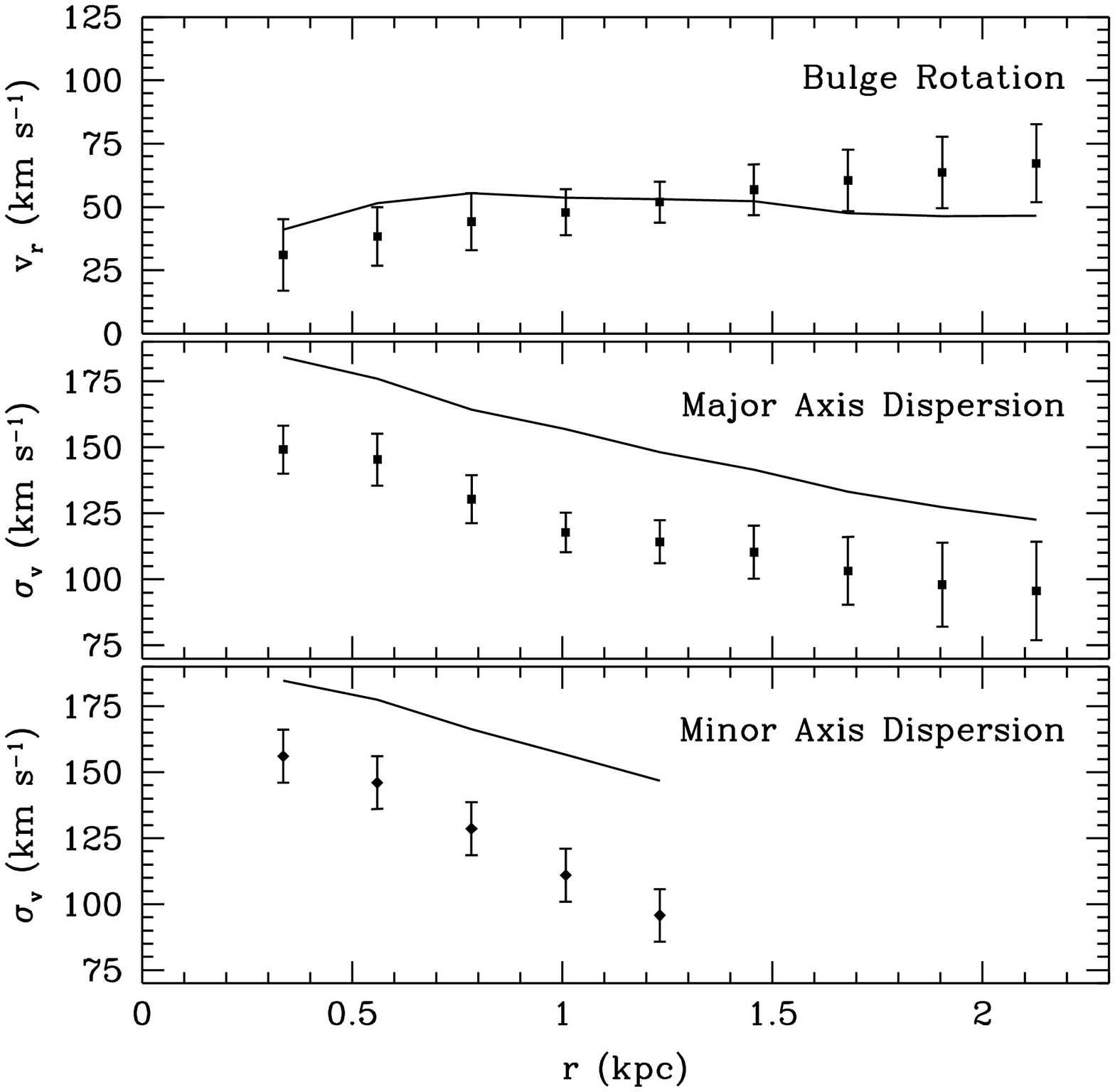}
\caption{Comparison of theoretical fits with observational data for
Model K1 \citep[][see Table~\protect\ref{tab:models}]{ken89}.  The
plots and line types are as described previously in
Figure~\protect\ref{fig:A}.}
\label{fig:K1}
\end{figure}

Model K1 assumes values for $M_d$ and $M_b$ from the popular
small-bulge model of \citet{ken89}.  This model does a reasonable job
of fitting the surface brightness profile as well as the rotation
curve beyond $6\,{\rm kpc}$ (Figure~\ref{fig:K1}).  However, Model K1
predicts a velocity dispersion in the bulge region that is too large.
This discrepancy is common among all models with $M_b\simeq 4\times
10^{10}\,M_\odot$ (e.g., Models C and K2).  It is already evident in
Figure 3 of \citet{ken89} if one focuses on the \citet{mce83} data at
$1-2\,{\rm kpc}$.  The discrepancy is worse for the self-consistent
models considered here.  Kent assumed a constant density halo: when
compared with a realistic model halo (i.e., one in which the density
is monotonically falling with radius) the contribution to the rotation
curve from a constant density halo is relatively low at small radii.
For a fixed halo contribution to $v_{\rm circ}^2$ at $30\,{\rm kpc}$,
Kent's model underestimates the halo contribution in the region of the
bulge as compared with more realistic models.

\begin{figure}
\plottwo{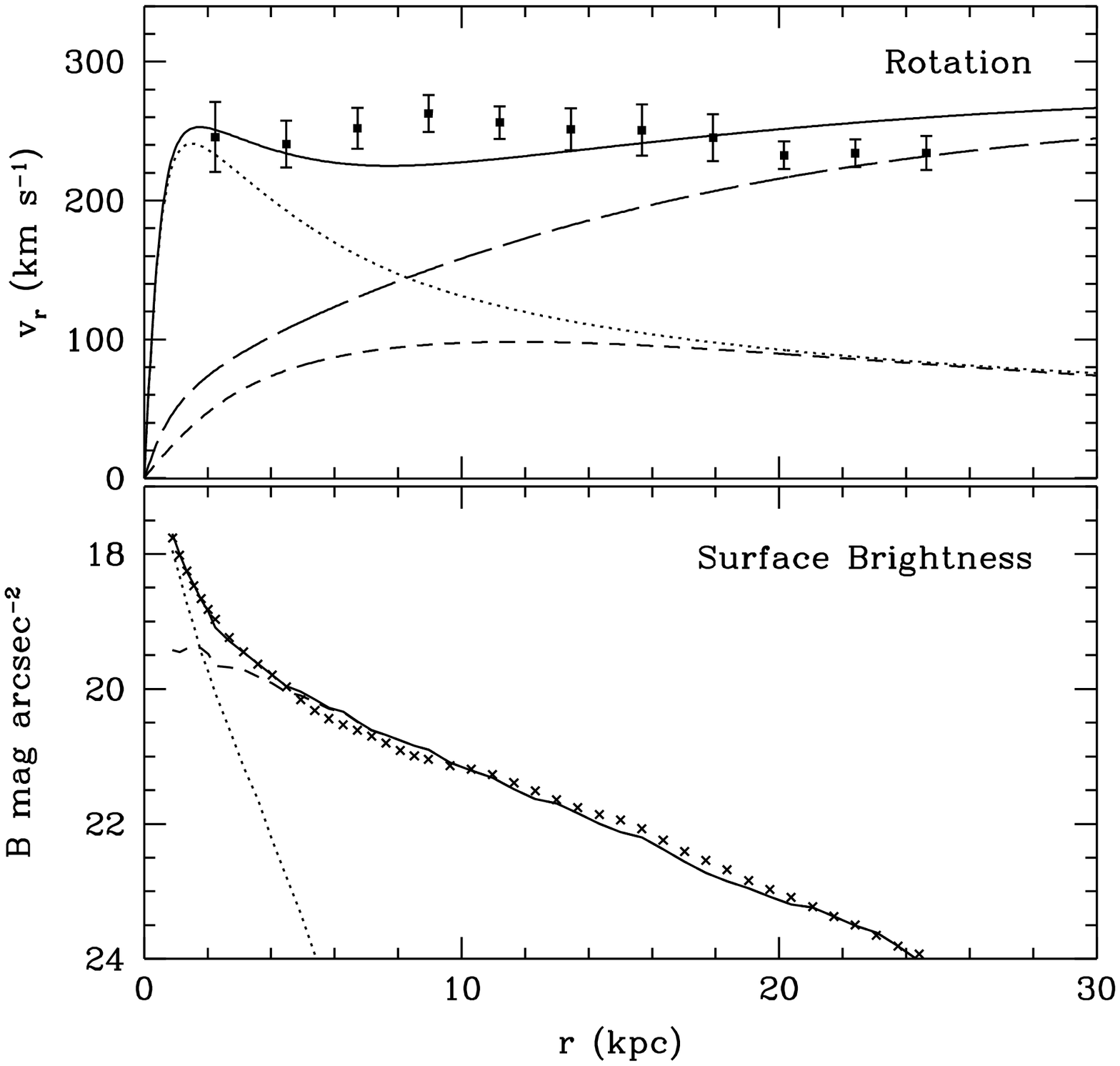}{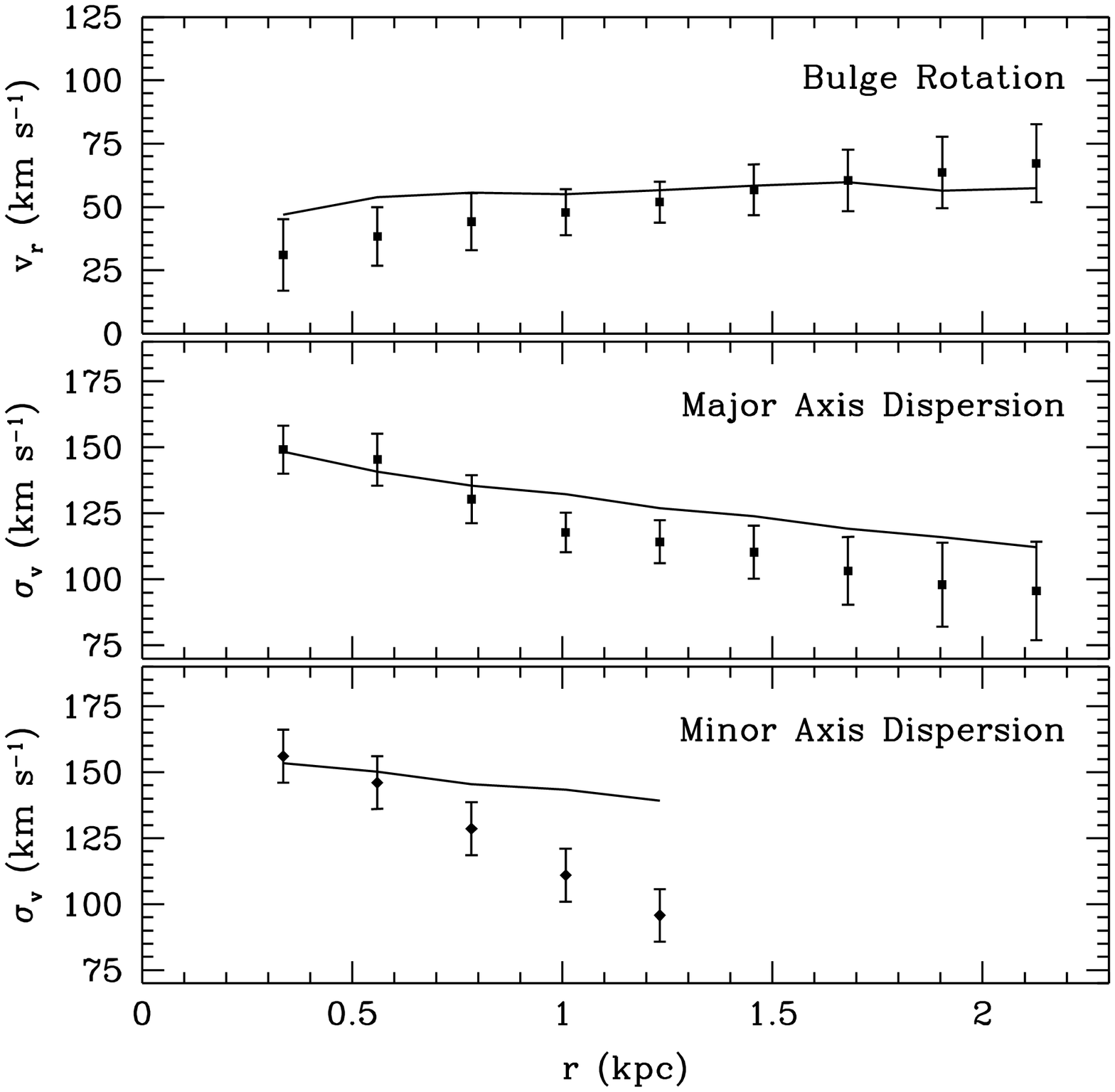}
\caption{Comparison of theoretical fits with observational data for
Model K2 (Values for $M_b$ and $M_d$ from \citet{ker01}.  See Table
~\protect\ref{tab:models}).  The
plots and line types are as described previously in
Figure~\protect\ref{fig:A}.}
\label{fig:K2}
\end{figure}

Model K2 assumes values for $M_d$ and $M_b$ used in the pixel-lensing
study from \citet{ker01}.  This model provides an excellent fit to the
surface brightness profile, matching almost perfectly the data through
the transition zone between bulge and disk dominated regions of the
galaxy (Figure \ref{fig:K2}).  However the model rotation curve
appears to have the wrong shape between $7$ and $12\,{\rm kpc}$ and
perhaps also beyond $20\,{\rm kpc}$.  In addition, the model velocity
dispersion profile is systematically high, as discussed above.
(Note that in \citet{ker01}, where the density profile of the halo
is taken to be that of a cored isothermal sphere, the model rotation
curve provides a good fit to the data.)

\subsection{Mass-to-Light Ratios}

We now consider the disk and bulge mass-to-light ratio for the models
in Table~\ref{tab:models} (see columns 5 and 6).  The $(M/L_R)$ values
for Model K1 are in excellent agreement with the results obtained by
\citet{ken89}, who found $\left (M/L_r\right )_d=10$ and $\left
(M/L_r\right )_b=5$.  This agreement serves as a consistency check of
our method though it is worth noting that Kent uses the r-band filter
of the Thuan \& Gunn system which differs slightly from the R-band
filter of the standard UBVRI system used by \citet{wal87} (see Table
2.1 of \citet{bin98} for a comparison of the different filter
characteristics).

As expected, the disk and bulge $M/L$ values for the other models in
Table \ref{tab:models} scale roughly with the mass of the
corresponding component.  In particular, the $M/L$ values for Model A
are approximately a factor of two smaller than those for Model K1.
The values for Model A compare well with the predictions from stellar
synthesis studies.  \citet{bel01} presented a correlation between the
stellar mass-to-light ratios and the optical colors of integrated
stellar populations for spiral galaxies.  Based on the relationships
they obtained under different assumptions of initial mass function
combined with typical M31 disk and bulge colors from Table~1 of
\citet{wal88}, one expects $M/L_R \sim 2-5$ within the different
regions of M31.  

Despite this general agreement between the model mass-to-light ratios
and the predictions from stellar population studies, the question
arises as to why, for all of the models of Table \ref{tab:models}
except K2, $\left (M/L_R\right )_d$ is significantly greater than
$\left (M/L_R\right )_b$.  In general, one expects the mass-to-light
ratio of the bulge to be comparable to, if not greater than, that of
the disk \citep[although see][]{her97}.  We propose two explanations
for this apparent difficulty.  First, the quoted $M/L_R$ values
presented in this paper do not incorporate corrections for foreground
extinction.  Since the effects of obscuration by dust are likely to be
more severe in M31's disk than in its bulge \citep{wal88}, this
correction would drive down the $M/L$ of the disk relative to that of
the bulge and bring the two values more into line \citep{ken89}.
Moreover, for typical Sb spirals like M31, gas contributes between 2
and 20\% of the mass of the disk \citep[see Figure 8.20 of][]{bin98}
and therefore boosts, by the same fraction, the $M/L$ value relative
to the value for the stellar population.  We therefore conclude that
the values for $\left (M/L\right )_d$ and $\left (M/L\right )_b$ are
quite reasonable.

It is nevertheless instructive to consider a second possibility,
namely that the $M/L$ values obtained for Model A reflect a genuine
problem that is perhaps be connected with the poor fit to the surface
brightness profile in the disk-bulge transition region
(Figure~\ref{fig:A}).  Deviations of the structure of M31's bulge from
a simple oblate spheroid may also have resulted in an underestimate of
its effective radius in the fitting procedure, causing the bulge to
get somewhat short-changed in mass.  One might then argue that Model
K2 provides a superior fit to M31.  However, the gains in the surface
brightness profile and the in mass-to-light ratios made by employing
Model K2 come at the cost of the quality of the rotation curve fit.
Model K2 also has the unusual feature that the bulge mass is greater
than the disk mass, contrary to what is commonly found for bright Sb
galaxies.


\subsection{Mass Distribution}

The total mass interior to a sphere $30\,{\rm kpc}$ in radius,
$M_{30}$, is provided in column 7 of Table~\ref{tab:models}.  These
values were obtained by generating an N-body realization of each model
and tabulating the mass interior to the prescribed sphere.  $M_{30}$
is constrained almost entirely by the outer rotation curve: a circular
rotation speed at $30\,{\rm kpc}$ of $v_r \simeq 215\,{\rm
km\,s^{-1}}$ (an average of the outer 4 points in
Figure~\ref{fig:rotation}) corresponds to a mass estimate $v_r^2 r/G =
32\times 10^{10}\,M_\odot$.  This is in good agreement with the values
obtained for most of our models.  The most prominent outlier is Model
K2, where the model rotation curve rises to $v_r\simeq 265\,{\rm
km\,s^{-1}}$ ($v_r^2 r/G \simeq 50\times 10^{10}\,M_\odot$), in
conflict with the observations.

Columns 8 and 9 of Table~\ref{tab:models} provide the halo mass and
tidal radius, respectively.  Since the data considered in this paper
probe only the inner $30\,{\rm kpc}$ of M31, it is not surprising that
$M_h$ and $R_t$ vary considerably.  Dynamical tracers such as globular
clusters and dwarf galaxy satellites have been used to constrain the
mass distribution of M31 beyond $r = 30\,{\rm kpc}$.  To better
understand how this type of data might help constrain the models, we
construct a sequence of models with \mbox{$80\,{\rm kpc}\la R_t\la
160\,{\rm kpc}$}, $q=1$, and $M_d$ and $M_b$ fixed to the values from
Model~A.  The results for Model~A ($R_t\simeq 80\,{\rm kpc}$) and
Model~E ($R_t\simeq 160\,{\rm kpc}$) are given in
Table~\ref{tab:Rmodels} for comparison.  In addition to $\chi^2$,
$R_t$, and $M_h$, we record the baryon fraction, $f_B$ (column 5).  We
also provide four quantities derived by substituting an NFW halo for
the lowered Evans model.  These quantities, to be discussed in
Section~\ref{sec:cosmology}, are the halo concentration parameter $c$
(column 6), the virial radius $R_{200}$ and mass $M_{200}$ (columns 7
and 8), and an alternate estimate for the baryon fraction based on the
NFW halo, $f_{B,NFW}$ (column 9).

\begin{figure}     
\plotone{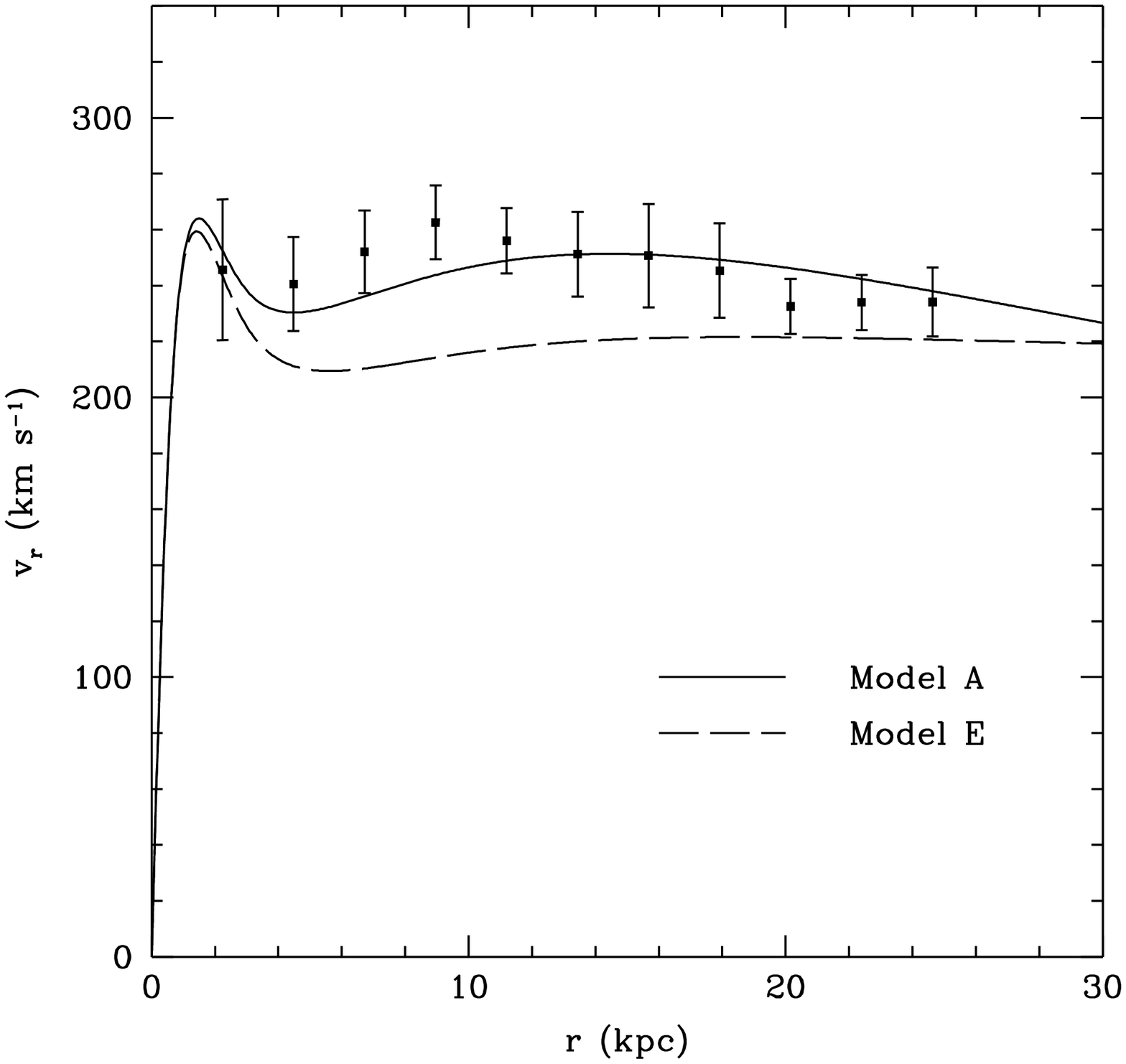}
\hspace*{\fill}\caption{Comparison of the rotation curves for 
Models A and E (see Table~\protect\ref{tab:Rmodels}).}
\label{fig:rotationR}  
\end{figure}

The value of $\chi^2$ is found to rise rapidly with increasing $R_t$.
The main source of the discrepancy is with the rotation curve fits, as
illustrated in Figure \ref{fig:rotationR}.  Models with $R_t>160\,{\rm
kpc}$ lead to even larger discrepancies between the predicted and
observed rotation curves and thus provide completely unacceptable fits
to the data.

Dynamical studies of globular clusters as well as planetary nebulae
are especially useful for obtaining mass estimates at intermediate
radii.  \citet{fed93} analyzed spectroscopic observations for several
dozen globular clusters in M31 between $10-30\,{\rm kpc}$, and derived
a range for $M_{30}$ of $(50-80)\times 10^{10}\,M_\odot$.  Using the
projected mass estimator of \citet{bah81} and \citet{hei85},
\citet{per02} obtained a mass estimate of $\sim 40\times
10^{10}\,M_\odot$ based on data from 319 globular clusters out to
$r\simeq 27\,{\rm kpc}$ and under the assumption of isotropic orbits.
As with most mass estimates that are based on dynamical tracers, the
uncertainty is due largely to our ignorance of the true orbit
distribution for the globular cluster system.

\citet{ew00} considered a sample of M31 globular cluster candidates
and planetary nebulae at large galactocentric radii and employed a
more sophisticated method for estimating the mass of the galaxy.  They
assumed simple analytic forms for the globular cluster DF and for the
halo-density-profile/gravitational-potential pair.  Two parameters
characterize their model potential and these were determined by
performing a maximum likelihood analysis over the data.  \citet{ew00}
obtained a mass estimate at $r= 40\,{\rm kpc}$ of $47\times
10^{10}\,M_\odot$ for the globular cluster data, while for the
planetary nebulae data they found $M_{30}\simeq 28\times
10^{10}\,M_\odot$.

At present, virtually all of the information available for the outer
halo of M31 comes from dynamical studies of its satellite galaxies.
\citet{cou99} used radial velocity data for Local Group members to
calculate a dynamical mass of the Andromeda subgroup of $(133 \pm
18)\times 10^{10}\,M_\odot$.  Using data from high-resolution
\'echelle spectroscopy from the Keck Telescope, \citet{cot00} applied
the projected mass estimator to derive an M31 mass of $\sim 79 \times
10^{10}\,M_\odot$ under the assumption of isotropic satellite orbits.
In the cases of circular and radial orbits, the estimated enclosed
masses change to $\sim 37\times 10^{10}\,M_\odot$ and $\sim 215\times
10^{10}\,M_\odot$, respectively.  Based on the dynamical modelling of
a similar set of data, \citet{eva00} obtained an M31 mass in the range
of $(70 - 100) \times 10^{10}\,M_\odot$.  For comparison, \citet{ew00}
derived an estimate for the total mass of M31 within $\sim 550\,{\rm
kpc}$ of $123^{+180}_{-60}\times 10^{10}\,M_\odot$.  This result was
obtained by applying the maximum likelihood DF method described above
to the combined globular cluster, planetary nebula, and satellite data
set.

\begin{figure}     
\plotone{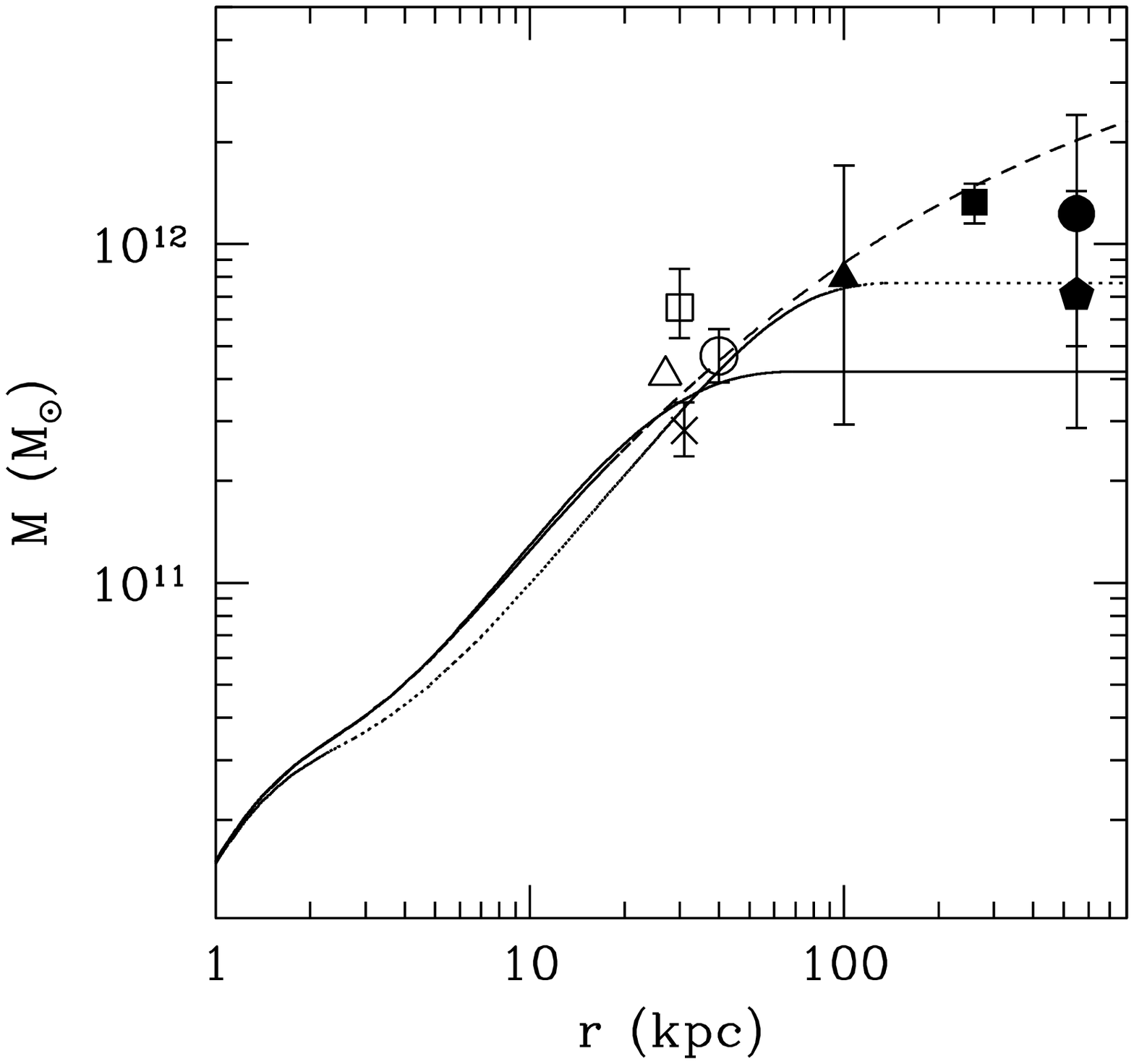}
\caption{Mass distribution for models of
Table~\protect\ref{tab:Rmodels}.  Plotted is the total mass interior
to the radius $r$ for Model A (solid line), Model E (dotted line),
and modified Model A with an NFW halo (dashed line).
Also shown are published mass estimates from studies of dynamical
tracers: globular cluster data from \citet[][open triangle]{per02},
\citet[][open circle]{ew00}, and \citet[][open square]{fed93};
satellite data from \citet[][filled triangle]{cot00}, \citet[][filled
pentagon]{eva00}, \citet[][filled circle]{ew00}, and \citet[][filled
square]{cou99}; planetary nebula data from \citet[][cross]{ew00}.}
\label{fig:mvsr}  
\end{figure}

In Figure~\ref{fig:mvsr} we plot the total mass distribution as a
function of radius for the models listed in Table~\ref{tab:Rmodels}.
The dynamical mass estimates described above are included for
comparison.  Based on these results, it would appear that both Models
A and E provide adequate descriptions of the outer halo, with Model E
coming closest to the estimates.

\begin{figure}     
\plotone{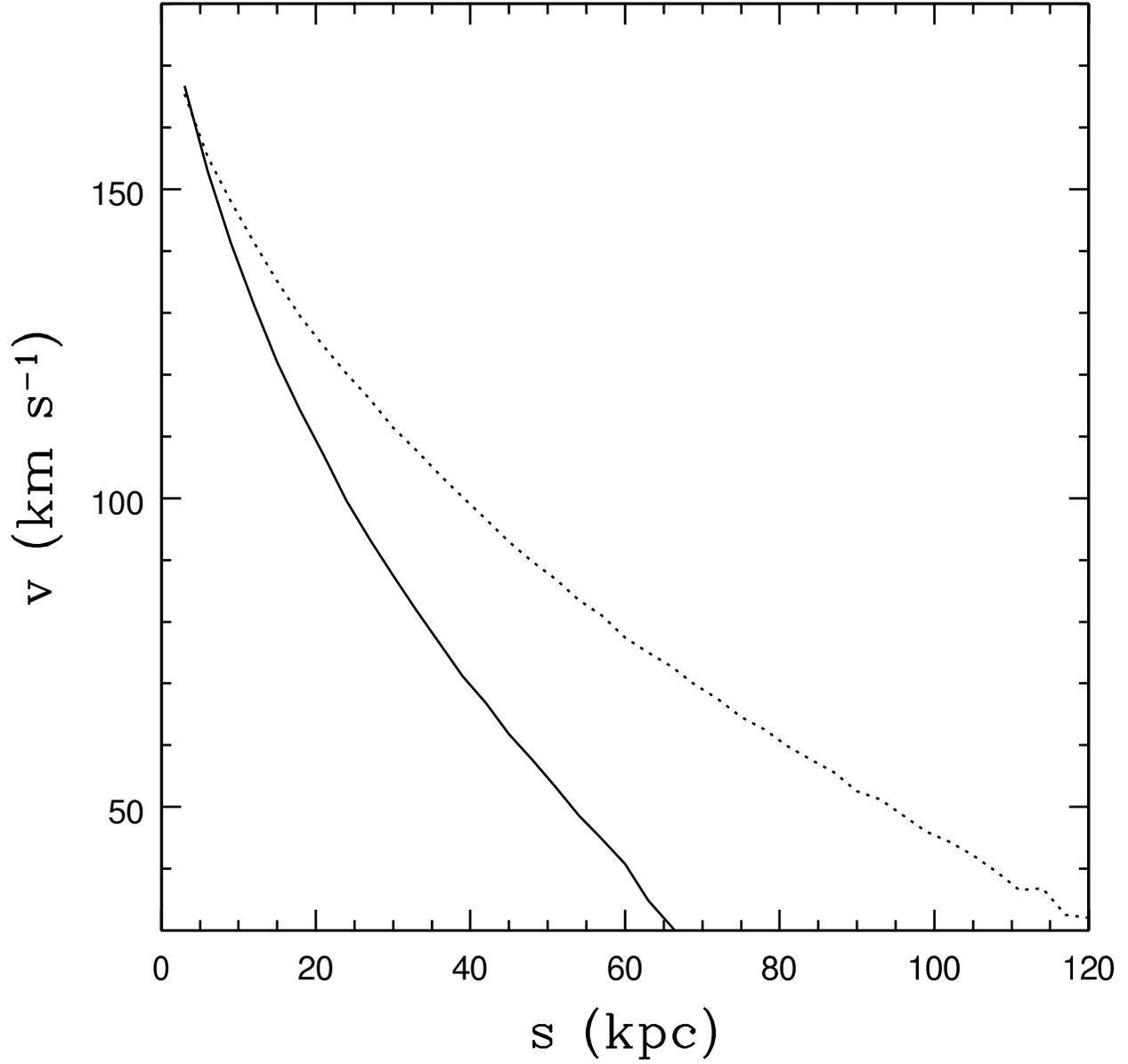}
\caption{Line-of-sight velocity dispersion as a function of projected
radius $s$.  Solid line refers to Model A; Dotted line refers to
Model E.}
\label{fig:los}  
\end{figure}

In a forthcoming publication, we will study in detail the constraints
on our models from observations of dynamical tracers.  Typically, for
each member of the tracer population, one knows its angular position
and line-of-sight velocity.  As an illustration of how one might use
such data, we show, in Figure~\ref{fig:los}, the line-of-sight
velocity dispersion as a function of projected radius for Models A and
E.  The velocity dispersion along a particular line of sight as a 
function of the projected position vector ${\bf s}$, is given by
\begin{equation}
\label{eq:los}
\sigma_{\rm LOS}^2\left ({\bf s}\right ) = 
\frac{1}{\Sigma\left ({\bf s}\right )}\int_{l_0}^\infty
dl\,\rho\left (l,{\bf s}\right )\langle v_l^2\left (l,{\bf s}\right )\rangle
\end{equation}
where $l_0$ corresponds to the position of the observer and
$\Sigma\left ({\bf s}\right )=\int_{l_0}^\infty dl\rho\left (l,{\bf
s}\right )$ is the surface density along the line of sight.  The
density $\rho\left (l,{\bf s}\right )$ and velocity dispersion
$\langle v_l^2\left (l,{\bf s}\right )\rangle$ are calculated from the
DF in the usual way:

\begin{equation}
\label{eq:rho}
\rho\left (l,{\bf s}\right ) = \int d^3 v \,f\left (l,{\bf s}\right )
\end{equation}
and
\begin{equation}
\label{eq:vel}
\rho\left (l,{\bf s}\right )
\langle v_l^2\left (l,{\bf s}\right )\rangle = 
\int d^3 v \,v_l^2\,f\left (l,{\bf s}\right ).
\end{equation}
Here $v_l$ is the component of the velocity along the line of sight.

KD provide a simple algorithm that allows one to generate an N-body
representation for any of their models.  An N-body representation can
be used to perform a Monte Carlo evaluation of the integrals in
Eqs.~\ref{eq:los}, \ref{eq:rho}, and \ref{eq:vel}.  The DF is written
as a sum over the particles:
\begin{equation}
\label{eq:nbody}
f\left ({\bf x},\,{\bf v}\right ) = m_i\sum_i
\delta^3\left ({\bf x}-{\bf x}_i\right )
\delta^3\left ({\bf v}-{\bf v}_i\right )
\end{equation}
where $m_i$, ${\bf x}_i$, and ${\bf v}_i$ are the mass, position, and
velocity of the $i^{\rm th}$ particle.  The surface density is given by

\begin{equation}
\label{eq:surface}
\Sigma\left ({\bf s}\right ) = \sum_{i\in{\cal V}} m_i
\end{equation}
where ${\cal V}$ is a volume corresponding to a thin tube centered on
the line of sight.  Likewise,

\begin{equation}
\label{eq:los2}
\sigma_{\rm LOS}^2\left ({\bf s}\right ) = 
\frac{1}{\Sigma\left ({\bf s}\right )}
\sum_{i\in{\cal V}} m_i v_{s,i}^2~.
\end{equation}
The curves in \ref{fig:los} represent an average over
position angle: differences between the line-of-sight dispersion
profiles along the major and minor axes were found to be
insignificant.  Our results may be compared with those from the
analytic halo-only model of \citet[][see their Figure 3]{eva00}.  More
to the point, the model prediction for the line-of-sight velocity
dispersion profile together with data for dynamical tracers can be
incorporated into an improved version of our model-finding algorithm.

As a final consistency check, we turn to the Tully-Fisher relation
\citep{tul77} which, in its original form, describes a tight correlation
between the total luminosity of a spiral galaxy and $v_{\rm flat}$,
the circular rotation speed in the flat part of the rotation curve.
For our purposes, a variant known as the baryonic Tully Fisher relation
\citep[see][]{mcg00}, which described a correlation
between $v_{\rm flat}$ and the total baryonic mass in gas and 
stars, $M_{\rm baryons}$ is more useful.  The baryonic Tully-Fisher
relation takes the form
\begin{equation}
M_{\rm baryons} = A\,v_{\rm flat}^b
\end{equation}
where $A$ and $b$ are constants.  \citet{mcg00} find that $b$ is
statistically indistinguishable from $4$ (however, see \citep{bel01})
and that if it is fixed to this value, the normalization is $A\simeq
35\,h_{75}^{-2}M_\odot {\rm km^{-4}s^4}$ with large uncertianties (see
for example, \citet{mcg01} where the acceptable range for $A$ is given
as $34-85$ in units of $M_\odot {\rm km^{-4}s^4}$).  For M31, $v_{\rm
(30\,kpc)} \approx 230$\,km\,s$^{-1}$ which implies, for a $A\simeq
35\,M_\odot {\rm km^{-4}s^4}$ baryonic mass of $9.8\times
10^{10}M_\odot$ in good agreement with the total baryon mass (taken to
be $M_d+M_b$) in Model A.  Note that somewhat higher values of $M_d$
can be tolerated without introducing too strong a bar instability.
Thus, one would be able to find a consistent model even if a higher
value of $A$ is assumed.

\subsection{Connection with Cosmology}
\label{sec:cosmology}

The baryon fraction can be used to constrain models of M31.  In
general, we expect the baryon fraction of a spiral galaxy such as M31
to be comparable to the baryon fraction of the Universe.  Under the
assumption that dark matter makes a negligible contribution to the
mass of the disk and bulge, we have \mbox{$f_B \le \left (M_d +
M_b\right )/ \left (M_d+M_b+M_h\right )$} where $f_B$ is the baryon
fraction of the galaxy as a whole.  The inequality incorporates the
possibility that some baryons may reside in the halo.  Values for
$f_B$ are given in column 8 of Table \ref{tab:Rmodels} and can be
compared with estimates from cosmology and astrophysics.  Based on Big
Bang nucleosynthesis constraints, \citet{bur01} derived an estimate of
$\Omega_Bh^2 = 0.020\pm 0.002$ where $\Omega_B$ is the density of
baryons in units of the critical density and $h$ is the Hubble
constant in units of $100\,{\rm km\,s^{-1}\,Mpc^{-1}}$.  Similar
results have been obtained from analyses of microwave background
anisotropy measurements.  For example, \citet{deB02} found
$\Omega_Bh^2 = 0.022^{+0.004}_{-0.003}$ for data from the BOOMERANG
experiment.  With values of $h=0.7\pm 0.07$ and $\Omega_{\rm matter} =
0.31\pm 0.13$ (the latter is also from \citet{deB02}) one finds $f_B =
0.13\pm 0.06$, the baryon fraction of the Universe.  This result is
consistent with estimates of the gas fraction in X-ray clusters
\citep{arm99}.

Figures \ref{fig:rotationR} and \ref{fig:mvsr} and the results for the
baryon fraction in Models A and E point to a potential problem with
using lowered Evans models for the halo of M31.  The baryon fraction
for Model A ($f_B = 0.23$) is too high.  Moreover, the total mass of
M31 in Model A falls in the lower range of acceptable values based on
the dynamics of satellites and globular clusters.  Model E does better
on both counts but its rotation curve provides a poor fit to the data.

These difficulties no doubt arise from the fact that the lowered Evans
models employed in the current implementation of the KD algorithm
incorporate a sharp cut-off in density at the tidal radius.  A natural
alternative is to use a model halo whose density profile falls off
more gradually with radius.  Such a substitution is consistent with
our current understanding of structure formation.  In the hierarchical
clustering scenario, each halo is a subsystem of a larger halo and in
this sense, the lowered Evans models are unphysical.

N-body simulations based on the cold dark matter model of structure
formation suggest that the density profiles of dark matter halos have
a simple universal shape.  This result was first noticed by
\citep[][hereafter NFW]{nfw96} who found that the spherically averaged
density profiles of halos in their simulations, which spanned four
orders of magnitude in mass, could be fitted by a function of the form
\begin{equation}
\label{eq:nfw}
\rho_{\rm NFW}(r) = \frac{\rho_s}{r/r_s\left (1 + r/r_s\right )^2}.
\end{equation}
Here $r_s$ and $\rho_s$ are free parameters which characterize the
radius and density in the transition region between the $r^{-1}$
central cusp and the $r^{-3}$ outer halo.  While there has been some
debate over the slope of the density profile at small radii, there is
widespread agreement that the NFW profile captures the general
features of dark matter halos in cosmological simulations.

It is common practice to define the virial radius as $R_{200}$, the
radius within which the mean density is $200$ times the background
density.  The concentration parameter, $c$, is defined as the ratio of
the virial radius to $r_s$: $c\equiv R_{200}/r_s$.  \citet{nfw96} (see
also \citet{bul01}) have found that there is a tight correlation
between $M_{200}$ (the mass interior to $R_{200}$) and $c$, the
implication being that the density profiles of simulated halos are
characterized by a single parameter.

In order to make contact with the results from cosmological
simulations, we determine the NFW profile that most closely matches
the halo profile of a particular KD model.  To be precise, we vary
$r_s$ and $\rho_s$ in Eq.\,\ref{eq:nfw} so as to minimize the RMS
deviation between the halo contribution to the rotation curve for the
KD model and the rotation curve derived from NFW (Eq.\,\ref{eq:nfw}).
Since our models were designed to fit rotation curve and velocity
dispersion data for $1\,{\rm kpc}<r<30\,{\rm kpc}$, only this range in
$r$ is used to determine the ``best-fit'' NFW model.

\begin{figure}     
\plotone{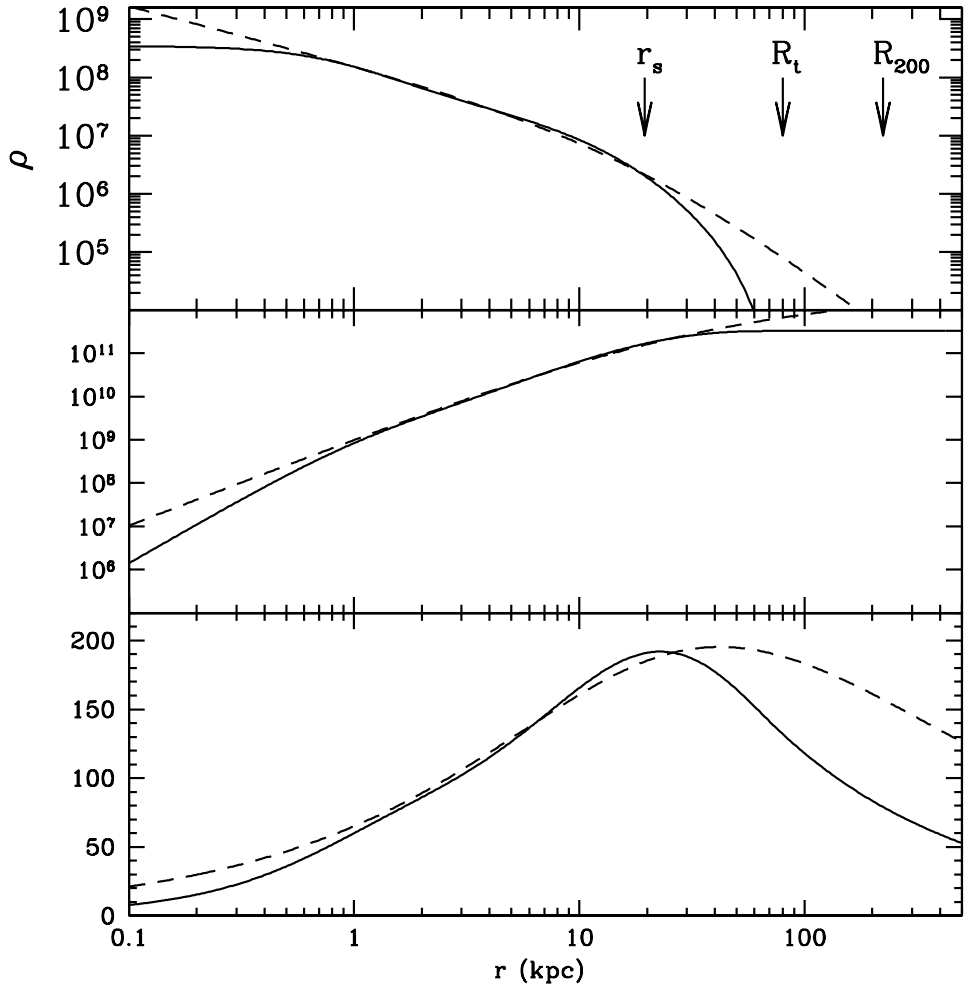}
\caption{Comparison of the halo used in Model A with a closely
matched NFW halo.  Top panel - density profile; Middle panel - mass
distribution; Bottom panel - halo contribution to the rotation curve.
Solid lines refer to Model A; dashed lines refer to the NFW halo.}
\label{fig:nfw}  
\end{figure}

We have carried out this exercise for Models A and E.  The values
obtained for $c$, $R_{200}$, and $M_{200}$ are given in columns 5-7 of
Table~\ref{tab:Rmodels}.  The density, mass distribution, and rotation
curve profiles are shown in Figure~\ref{fig:nfw}.  We see that for the
range in radius probed by the data considered in this paper, the
rotation curve of an NFW model can be matched closely to the halo
contribution to the rotation curve for our models.  Moreover, the
values obtained for $c$ are consistent with what is predicted by the
cosmological simulations \citep{nfw96, bul01}.

Not surprisingly, the halo profiles for the KD models differ
significantly from the NFW profile at small and large radii.  The
lowered Evans models have a constant density core whereas the NFW
models have an $r^{-1}$ cusp.  The halos in the self-consistent models
develop a weak cusp but the slope is closer to $0$ than to $-1$.
However, the discrepancy in mass at small radii between the two models
is actually quite small: at $r=1\,{\rm kpc}$, the cumulative mass for
the NFW model exceeds that of the KD model by an amount $\la
10^8\,M_\odot$.

The discrepancy at large radii is more interesting.  The lowered Evans
models have a sharp cut-off in density whereas the NFW profile falls
off as $r^{-3}$.  The discrepancy at large radii is especially
apparent in Model A where $R_{200}$ is nearly a factor of $3$ larger
than $R_t$ and $M_{200}$ is a factor of $4$ larger than $M_h$.

Figure~\ref{fig:nfw} suggests a resolution to the baryon fraction
problem discussed above, namely to replace the halo in Model A with an
NFW profile selected to match the rotation curve within $30\,{\rm
kpc}$.  The excellent overall fit to the data is retained and the
baryon fraction is brought down to a value comfortably within the
range predicted by cosmology.  At this stage, a cautionary remark is
in order.  The KD models, by design, describe dynamically
self-consistent systems and one cannot simply substitute a different
halo with an NFW profile without spoiling the self-consistency.
Several authors have derived DFs for NFW models \citep[e.g.,][]{zha97,
wid00, lok00} which, with some modification, could be incorporated
into future implementations of the KD algorithm.

\subsection{Halo Shape}

Our final set of models explores variations in the shape of the dark
halo.  Results for two of these models are provided in Table
\ref{tab:Qmodels}.  For Model Q1, we set $q=0.85$ and leave $R_t$
unconstrained, whereas for Model Q2 the target value of $R_t$ is set
to be $60\,{\rm kpc}$.  The results are summarized in Table 2 where as
before we present $\chi^2$ (column 3), $R_t$ (column 4), and $M_h$
(column 5).  We also provide $q_h$, an effective flattening parameter
for the mass distribution (column 6).  The value of $q_h$ is
calculated by taking the ratio of the mass moments along the short and
long axes.  It differs from $q$ for two reasons.  First, the halo
``responds'' to the disk through the Poisson-solving algorithm of KD.
Thus, even for $q=1$, the mass distribution is flattened slightly and
$q_h< 1$.  Second, $q_h$ reflects the mass distribution which tends to
be more aspherical than the gravitational potential.  Therefore,
$q_h<q$ for models Q1 and Q2.

\begin{figure}     
\plotone{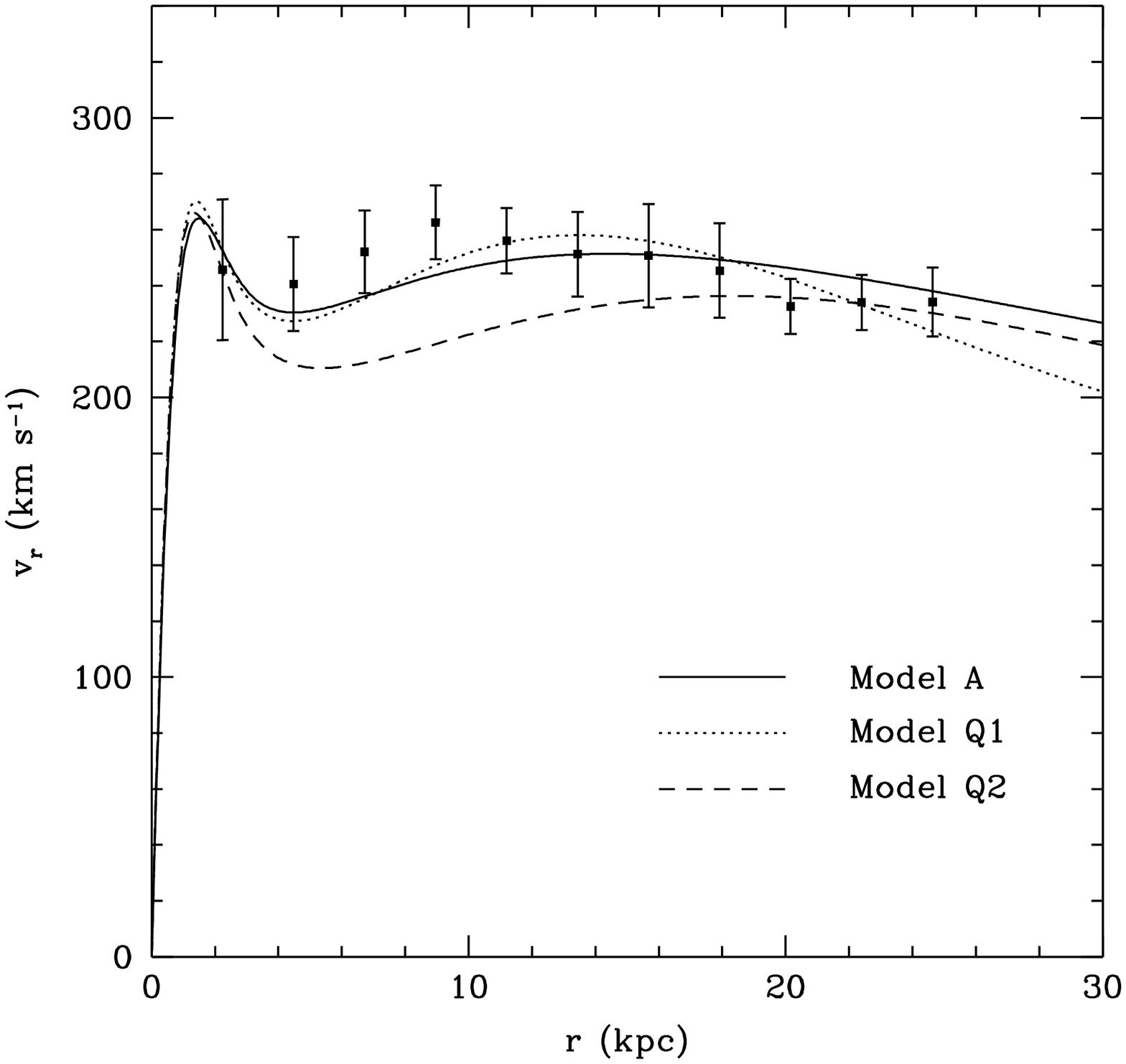}
\caption{Comparison of the rotation curves for Models A, Q1 and Q2
(see Table~\protect\ref{tab:Qmodels}).}
\label{fig:rotationQ}  
\end{figure}

Flattened models appear to favor smaller tidal radii.  Model Q1, for 
example, provides an excellent fit to the data but assumes a tidal radius
that is very small.  This model is probably inconsistent with observations
of satellites beyond $30\,{\rm kpc}$.  Model Q2 has a somewhat larger
tidal radius but, as with Model E, the fit to the rotation curve
is significantly degraded (Figure \ref{fig:rotationQ}).

\section{Gravitational Microlensing}

Gravitational microlensing is an attractive means by which to detect
MACHOs in the halo of our galaxy and the halos of our nearest
neighbors.  Results from the 5.7-year LMC data set of the MACHO
collaboration suggest that a significant new component of the Galaxy
--- namely one composed of MACHOs --- has been discovered
\citep{alc00}.  The MACHO collaboration detected 13-17 microlensing
events, a number too large to be accounted for by known populations.
Within the context of a specific halo model, a maximum likelihood
analysis yields an estimate for the MACHO halo fraction of $20\%$ with
the most likely MACHO mass to be between $0.15$ and $0.9\,M_\odot$.
This result is puzzling since it requires rather extreme assumptions
about star formation and galaxy formation.  Moreover, the result is in
conflict with the upper limits on the halo mass fraction found by the
EROS collaboration \citep{las00}.  One possibility is that the lenses
responsible for the observed microlensing events are in the Magellanic
clouds or in the Galactic disk rather than the Galactic halo.

Microlensing surveys toward M31 have the potential to resolve this
question.  The main advantage of looking to M31 is that one can probe
a variety of lines of sight across the M31 disk and bulge and through
its halo.  In particular, a massive spherical halo of MACHOs will
yield more events toward the far side of the disk than toward the near
side.  This front-back asymmetry is an unambiguous signature of a
MACHO halo \citep{cro92}.

A number of authors have computed theoretical event rate maps
\citep{gyu00,ker01,bal02} assuming an ad hoc model for the disk,
bulge, and halo of M31.  Only cursory attempts were made to insure
that the models are dynamically self-consistent.  In this regard, our
models represent an improvement over the models considered in the
aforementioned papers.

\begin{figure}     
\plotone{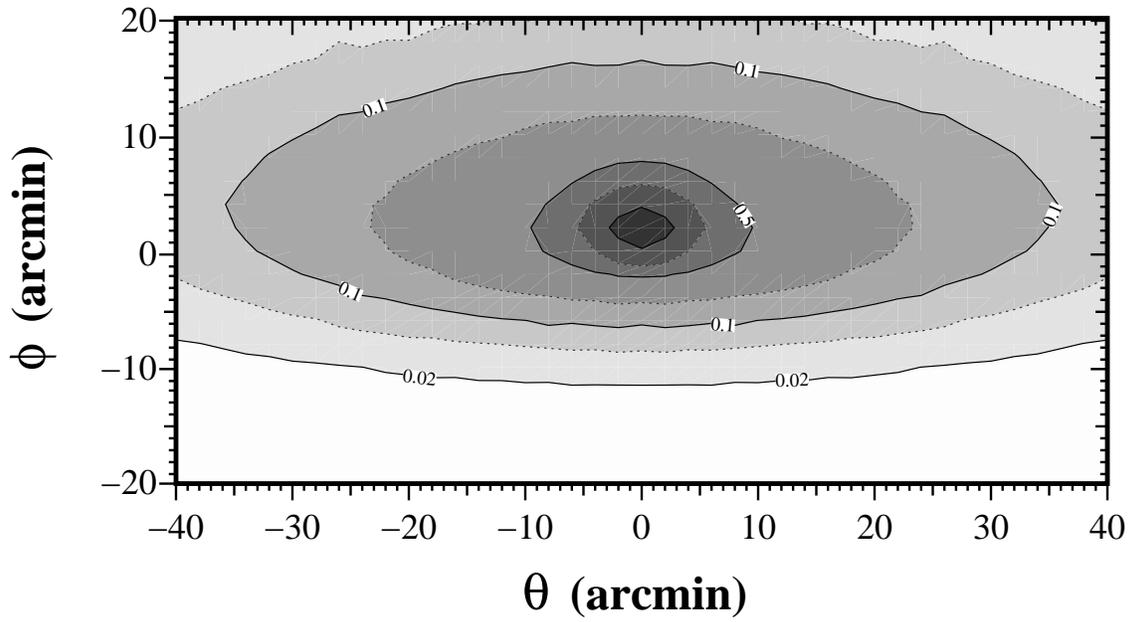}
\caption{Contours of concurrent microlensing events per arcmin$^2$ 
for Model A assuming an all MACHO halo.  Contours are, from
the outside in, 0.02, 0.05, .1, .2, .5 1.0, 2.0.}
\label{fig:optical_A}  
\end{figure}

Following \citet{gyu00} we consider the quantity $d\tau/dA$, the
number of concurrent events per area on the sky.  This quantity is
roughly the product of the optical depth (number of concurrent events
per source star) and the surface density of sources, and is more
closely aligned with what the experiments measure than the optical
depth.  As with the line-of-sight velocity dispersion, the N-body
representation of the KD models allows one to calculate theoretical
optical depth and event rate maps quickly and efficiently.  The
quantity $d\tau/dA$ is evaluated by performing a double integral over
source and lens distributions:

\begin{equation}
\label{eq:optical}
\frac{d\tau}{dA} = \int_0^\infty dL n_{\rm source}(L)
\int_0^L dl\,\frac{\rho_{\rm lens}(l)}{M_{\rm lens}}\,\frac{4GM_{\rm lens}}{c^2}\,\frac{L-l}{L}
\end{equation}
where $n_{\rm source}$ is the number density of sources a distance $L$
from the observer, $\rho_{\rm lens}$ is the mass density in lenses a
distance $l$ from the observer, and $M_{\rm lens}$ is the mass of the
lens.  Given an N-body representation of a galaxy, this integral may
be evaluated by performing the double sum:

\begin{equation}
\label{eq:nbody_optical}
\frac{d\tau}{dA} = \sum_{i\in\rm source} 
~\sum_{j\in\rm lens} \,
\frac{4GM_{\rm lens}}{c^2}\,\frac{L_i-l_j}{L_i}
\end{equation}
The map of the number of concurrent events per arcmin$^2$ for Model A
is shown in Figure \ref{fig:optical_A}.

The differential event rate (number of events per unit time as a
function of duration and position across the M31 disk) may be
calculated in a similar manner.  A semi-analytic calculation, on the
other hand, involves multidimensional integrals which may be
prohibitively complicated depending on the functional form of the DFs.


\section{Summary and Conclusions} 

The models for M31 presented in this paper are dynamically
self-consistent, consistent with published observations, and stable
against the rapid growth of bar-like modes in the disk.  To the best
of our knowledge, no other model of a disk galaxy satisfies these
three criteria to the extent considered in this paper.  Although KD
constructed models for the Milky-Way that are dynamically
self-consistent, stable, and have roughly the correct rotation curve,
they did not attempt to fit their model to other types of data such as
the surface brightness profile.

Our models span a wide range in halo size and shape.  The most
successful models assume a bulge mass that is nearly a factor of two
smaller than the oft-quoted value from \citet{ken89}.  Our galaxy
model with $M_b = 2.5\times 10^{10}\,M_\odot$ and $M_d = 7\times
10^{10}\,M_\odot$ provides a good overall fit to observational data,
yields mass-to-light ratios that are quite acceptable, and appears to
be stable against bar formation.  This disruptive bar instability
becomes more apparent in the models with more massive disks, such as
the model proposed by \citet{ken89}.  At the other extreme lies the
model of \citet{ker01}, having $M_b = 4\times 10^{10}\,M_\odot$ and
$M_d = 3\times 10^{10}\,M_\odot$.  Although their model is extremely
stable against bar formation and reproduces the surface brightness
profile nicely within the disk-bulge transition zone, the M31 rotation
curve and inner velocity dispersion profile it provides do not fit the
data at all well.

The favored model in our study is found to yield a poor match of the
estimated baryon fraction with cosmological predictions, and it falls
somewhat short of the total galaxy mass at large radii as determined
from observations of dynamical tracers.  Forcing a larger tidal radius
($R_t \sim 160$\,kpc) improves these at the cost of the rotation curve
fit.  We attribute this failure to the form of the lowered Evans halo
DF utilized in the KD algorithm, and propose the use of a halo DF that
falls off more gradually with radius.  A preliminary analysis suggests
that if the halo of Model A is replaced by an appropriate NFW profile,
the cosmological constraints are satisfied while the quality of the
fit to the observational data is maintained.

Several improvements in the models are possible.  For example, one can
incorporate additional components of the galaxy such as a central
black hole, thick disk, and stellar halo.  The most serious drawback
of the models is that they are axisymmetric and therefore cannot
capture important aspects of M31 such as spiral structure and
triaxiality of the bulge.  Our models can serve as a starting point
for investigations of these phenomena.  In particular, models with
$M_d\simeq 7\times 10^{10}\,{\rm M_\odot}$ will be studies using
numerical simulations to see if we can produce spiral structure and a
barlike bulge similar to what is observed in M31.

The methods described in this paper are completely general.  When
additional data become available, the algorithm can be rerun to
determine a new suite of best-fit models.  One extension that is soon
to be implemented is the inclusion of a distribution of test-particles
that are designed to represent a population of dynamical tracers such
as globular clusters or satellite galaxies.  In principle, the
additional information from observations of these populations can
constrain the extent and shape of the halo.


\acknowledgements{We are grateful to J. Dubinski and K. Kuijken for
providing us with their code and for invaluable assistance in running
it.  We also thank E. Baltz, A. Crotts, G. Gyuk, J. Irwin, S. Kent,
and D. Stiff for useful conversations and to J. Dubinski for comments
and suggestions based on an early draft of this manuscript.  
We also thank the anonymous referee for useful suggestions.  This work
was supported, in part, by the Natural Science and Engineering
Research Council of Canada.}

\clearpage
\appendix

\section{KD model parameters}

The input parameters for the KD algorithm are provided in
Table~\ref{tab:KDpars} for the models presented in this paper.  The
parameters are as discussed in the text.  The unit of length is
$1\,{\rm kpc}$ and the unit of velocity is $100\,{\rm km\,s^{-1}}$.
We set $G_{\rm newt}=1$, as is the convention in N-body simulations.
This implies of unit of mass of $2.325\times 10^{-9}\,M_\odot$.




\newpage

\begin{deluxetable}{ccccccccc}
\tablewidth{0pt}
\tabletypesize{\footnotesize}
\tablecaption{Models with $q=1$ and $R_t$ unconstrained\label{tab:models}}
\tablehead{
 \colhead{Model} & \colhead{$M_d$} & \colhead{$M_b$} & \colhead{$\chi^2$} &
 \colhead{$\left (M/L_R\right )_d$} & \colhead{$\left (M/L_R\right )_b$} & 
 \colhead{$M_{30}$} & \colhead{$M_h$} & \colhead{$R_t$} \\
 \colhead{} & \colhead{($10^{10}\,M_\odot$)} & \colhead{($10^{10}\,M_\odot$)} &
 \colhead{} & \colhead{} & \colhead{} & \colhead{($10^{10}\,M_\odot$)} & 
 \colhead{($10^{10}\,M_\odot$)} & \colhead{(kpc)} 
}
\startdata
A  & 7  & 2.5 & 0.70 & 4.4 & 2.7 & 36  & 32   &  80   \\
B  & 7  & 1   & 2.29 & 4.5 & 1.3 & 38  & 27   &  38   \\
C  & 7  & 4   & 1.55 & 4.4 & 4.1 & 29  & 33   & 129   \\
D  & 14 & 2   & 0.63 & 8.6 & 2.4 & 32  & 32   & 201   \\
K1 & 16 & 4   & 1.75 & 9.9 & 4.8 & 28  & 6.5  & 137   \\ 
K2 & 3  & 4   & 1.28 & 2.0 & 3.2 & 50  & 120  & 155   \\ 
\enddata
\end{deluxetable}

\begin{deluxetable}{ccccccccc}
\tablewidth{0pt} 
\tablecaption{Models with $M_d = 7\times 10^{10}\,M_\odot$,
$M_b = 2.5\times 10^{10}\,M_\odot$, and $q=1$
\label{tab:Rmodels}}
\tablehead{ 
 \colhead{Model} & \colhead{$R_t$} &
 \colhead{$\chi^2$} & \colhead{$M_h$} & \colhead{$c$} &
 \colhead{$R_{200}$} & \colhead{$M_{200}$} & \colhead{$f_B$} &
 \colhead{$f_{B,NFW}$}\\
 \colhead{} & \colhead{(kpc)} & \colhead{} & 
 \colhead{($10^{10}\,M_\odot$)} & 
 \colhead{} & \colhead{(kpc)} & \colhead{($10^{10}\,M_\odot$)} &
 \colhead{} & \colhead{}}
\startdata
A &  80 & 0.70 & 32 & 11.5 &  224 & 130 & 0.23 & 0.07 \\
E & 160 & 1.30 & 67 &  7.9 &  193 & 81  & 0.13 & 0.10 \\
\enddata
\end{deluxetable}

\begin{deluxetable}{cccccc} 
\tablewidth{0pt}
\tablecaption{Models with $M_d = 7\times 10^{10}\,M_\odot$, 
$M_b = 2.5\times 10^{10}\,M_\odot$, and $R_t$ unconstrained\label{tab:Qmodels}}
\tablehead{
 \colhead{Model} & \colhead{$q$} &\colhead{$\chi^2$} & 
 \colhead{$R_t$ (kpc)} & \colhead{$M_h$ ($10^{10}\,M_\odot$)} &
 \colhead{$q_h$} 
}
\startdata
A  & 1.00 & 0.70 &  80 & 32 &  0.93  \\ 
Q1 & 0.85 & 0.63 &  37 & 17 &  0.59  \\ 
Q2 & 0.85 & 1.08 &  58 & 23 &  0.61  \\ 
\enddata
\end{deluxetable}

\begin{deluxetable}{ccccccccccc}
\tablewidth{0pt}
\tablecaption{Input KD parameters for models in 
Tables~\protect\ref{tab:models}, \protect\ref{tab:Rmodels}, 
and \protect\ref{tab:Qmodels}.\label{tab:KDpars}}
\tablehead{
\colhead{Model} & 
\colhead{$\Psi_0$} & \colhead{$\sigma_0$} & \colhead{$q$} & 
  \colhead{$(\frac{r_c}{r_K})^2$} & \colhead{$R_a$} & \colhead{$m_{disk}$} & 
\colhead{$\rho_b$} & \colhead{$\Psi_{c}$} & \colhead{$\sigma_b$} & 
  \colhead{$S_b$} 
}
\startdata
A & -28.87 & 4.20 & 1.00 & 0.43 & 4.77 & 30.11 & 6.68 & -16.26 & 2.24 & 0.80\\
B & -24.52 & 4.27 & 1.00 & 0.41 & 4.73 & 30.11 & 7.07 & -18.55 & 2.24 & 0.85\\
C & -32.29 & 4.34 & 1.00 & 0.40 & 5.14 & 30.11 & 6.52 & -14.79 & 2.41 & 0.81\\
D & -29.77 & 3.37 & 1.00 & 0.42 & 4.50 & 60.22 & 6.87 & -17.96 & 2.05 & 0.77\\
E & -28.98 & 3.68 & 1.00 & 0.40 & 6.03 & 30.11 & 7.35 & -16.68 & 2.26 & 0.81\\
K1 & -34.86 & 3.58 & 1.00 & 0.43 & 4.30 & 68.82 & 6.38 & -16.03 & 2.44 & 0.75\\
K2 & -35.21 & 3.62 & 1.00 & 0.15 & 5.26 & 12.90 & 4.94 & -17.90 & 1.62 & 0.82\\
Q1 & -26.57 & 4.05 & 0.85 & 0.42 & 3.63 & 30.11 & 9.70 & -13.81 & 2.57 & 0.80\\
Q2 & -26.23 & 3.79 & 0.85 & 0.44 & 4.27 & 30.11 & 10.18 & -13.74 & 2.55 & 0.80\\
\enddata
\end{deluxetable}


\begin{thebibliography}{}

\bibitem[Alcock et al.(2000)]{alc00} Alcock, C., et al. 2000,
 \apj, 542, 281
\bibitem[Armaud \& Evrard(1999)]{arm99} Armaud, M., \& Evrard, A. E.
 1999, \mnras, 305, 631
\bibitem[Bahcall \& Tremaine(1981)]{bah81} Bahcall, J. N., \& Tremaine, S.
 1981, \apj, 244, 805
\bibitem[Baltz, Gyuk \& Crotts(2002)]{bal02} Baltz, E. A., Gyuk, G.,
 \& Crotts, A. 2002, astro-ph/0201054
\bibitem[Barnes \& Efstathiou(1987)]{bar87} Barnes, J., \& Efstathiou,
 G. 1987, \apj, 319, 575
\bibitem[Bell \& de Jong(2001)]{bel01} Bell, E. F., \& de Jong, R. S. 2001,
\apj, 550, 212
\bibitem[Binney \& Merrifield(1998)]{bin98} Binney, J., \& Merrifield, M. 1998,
 Galactic Astronomy, Princeton Univ. Press, Princeton
\bibitem[Binney \& Tremaine(1987)]{bin87} Binney, J., \& Tremaine, S. 1987,
 Galactic Dynamics, Princeton Univ. Press, Princeton
\bibitem[Braun(1991)]{bra91} Braun, R. 1991, \apj, 372, 54
\bibitem[Bullock et al.(2001)]{bul01} Bullock, J. S., Kolatt, T. S.,
 Sigad, Y., Somerville, R. S., Kravtsov, A. V., Klypin, A. A.,
 Primack, J. R., \& Dekel, A. 2001, \mnras, 321, 559
\bibitem[Burles, Nollett, \& Turner(2001)]{bur01} Burles, S., Nollett, K. N.,
 \& Turner, M. S. 2001, \apj, 552, L1
\bibitem[Byrd(1978)]{byr78} Byrd, G. G. 1978, \apj, 226, 70 
\bibitem[Calchi Novati et al.(2002)]{cal02} Calchi Novati, S., et al. 2002,
\aap, 381, 848
\bibitem[C\^ot\'e et al.(2000)]{cot00} C\^ot\'e, P., Mateo, M.,
 Sargent, W. L. W., \& Olszewski, E. W. 2000, \apj, 537, L91
\bibitem[Courteau \& van den Bergh(1999)]{cou99} Courteau, S., \& van
den Bergh, S. 1999, \aj, 118, 337
\bibitem[Cram, Roberts \& Whitehurst(1980)]{cra80} Cram, T. R., Roberts, 
  M. S., \& Whitehurst, R. N. 1980, \aaps, 40, 215
\bibitem[Crotts(1992)]{cro92} Crotts, A. P. S. 1992, \apj, 399, L43
\bibitem[Crotts \& Tomaney(1996)]{cro96} Crotts, A., \& Tomaney, A. 1996,
 \apj, 473, L87
\bibitem[Crotts et al.(2001)]{cro01} Crotts, A., Uglesich, R.,
 Gould, A., Gyuk, G., Sackett, P., Kuijken, K., Sutherland, W., \& 
 Widrow, L. 2000 in Microlensing 2000: A New Era of Microlensing Astrophysics,
 ASP Conf. Proc. 239, ed. J. W. Menzies \& P. D. Sackett
\bibitem[de Bernardis et al.(2002)]{deB02} de Bernardis, P., et al. 2002, 
\apj, 564, 559
\bibitem[Deharveng \& Pellet(1975)]{deh75} Deharveng, J. M., \&
  Pellet, A. 1975, \aap, 38, 15
\bibitem[Dehnen(2002)]{deh02} Dehnen, W. 2002, J. Comp. Phys., 179, 27
\bibitem[Dubinski(1994)]{dub94} Dubinski, J. 1994, \apj, 431, 617
\bibitem[Evans(1993)]{eva93} Evans, N. W. 1993, \mnras, 260, 191
\bibitem[Evans \& Wilkinson(2000)]{ew00} Evans, N. W., \& Wilkinson, M. I.
 2000, \mnras, 316, 929
\bibitem[Evans et al.(2000)]{eva00} Evans, N. W., Wilkinson, M. I.,
 Guhathakurta, P., Grebel, E. K., \& Vogt, S. S. 2000, \apj, 540, L9
\bibitem[Evans et al.(2002)]{eva02} Evans, N. W., Wilkinson, M. I., 
  Perrett, K. M., \& Bridges, T. J. 2002, \apj, in press
\bibitem[Federici et al.(1993)]{fed93} Federici, L., B\`onoli, F., Ciotti, L., 
 Fusi-Pecci, F., Marano, B., Lipovetsky, V. A., Niezvestny, S. I., 
 \& Spassova, N. 1993, \aap, 274, 87
\bibitem[Gerhard(1986)]{ger86} Gerhard, O. E. 1986, \mnras, 219, 373
\bibitem[Gottesman \& Davies(1970)]{got70} Gottesman, S. T., \& Davies, R. D.
  1970, \mnras, 149, 263
\bibitem[Gyuk \& Crotts(2000)]{gyu00} Gyuk, G., \& Crotts, A. 2000,
 \apj, 535, 621
\bibitem[Heisler, Tremaine \& Bahcall(1985)]{hei85} Heisler, J.,
Tremaine, S., \& Bahcall, J. N. 1985, \apj, 298, 8
\bibitem[H\'eraudeau \& Simien(1997)]{her97} H\'eraudeau, P., \&
Simien, F. 1997, \aap, 326, 897
\bibitem[Hodge \& Kennicutt(1982)]{hod82} Hodge, P. W., \& Kennicutt,
  R. C. 1982, \aj, 87, 264
\bibitem[Hohl(1971)]{hoh71} Hohl, F. 1971, \apj, 168, 343
\bibitem[Kent(1989)]{ken89} Kent, S. 1989, \pasp, 101, 489
\bibitem[Kerins et al.(2001)]{ker01} Kerins, E., et al. 2001, \mnras, 323, 13
\bibitem[King(1966)]{kin66} King, I. R. 1966, \aj, 71, 64
\bibitem[Kuijken \& Dubinski(1994)]{kui94} Kuijken, K., \& Dubinski, J.
  1994, \mnras, 269, 13
\bibitem[Kuijken \& Dubinski(1995)]{kui95} Kuijken, K., \& Dubinski, J.
  1995, \mnras, 277, 1341 (KD)
\bibitem[Lokas \& Mamon(2000)]{lok00} Lokas, E. L., \& Mamon, G. A. 2000,
\mnras, 321, 155
\bibitem[Lasserre et al.(2000)]{las00} Lasserre, T. et al. (2000), \aap, 
 355, L39
\bibitem[Mateo(1998)]{mat98} Mateo, M. 1998, \araa, 36, 435
\bibitem[McElroy(1983)]{mce83} McElroy, D. B. 1983, \apj, 270, 485
\bibitem[McGaugh(2001)]{mcg01} McGaugh, S. 2001, astro-ph/0112357
\bibitem[McGaugh et al.(2000)]{mcg00} McGaugh, S. S., Schombert, J. M., 
Bothun, G. D., \& de Blok, W. J. G. 2000, \apj, 533, L99 
\bibitem[Navarro, Frenk \& White(1996)]{nfw96} Navarro, J. F.,
 Frenk, C. S., \& White, S. D. M. 1996, \apj, 462, 563
\bibitem[Nolthenius \& Ford(1987)]{nol87} Nolthenius, R., \& Ford, 
 H. C., 1987, \apj, 317, 62
\bibitem[Ostriker \& Peebles(1973)]{ost73} Ostriker, J. P., \&
Peebles, P. J. E.  1973, \apj, 186, 467
\bibitem[Perrett et al.(2002)]{per02} Perrett, K. M., Bridges, T. J., 
  Hanes, D. A., Irwin, M. J., Brodie, J. P., Carter, D., Huchra, J. P., 
  \& Watson, F. G. 2002, \aj, 123, 2490 
\bibitem[Press et al.(1986)]{pre86} Press, W. H., Flannery, B. P.,
  Teukolksy, S. A., \& Vetterling, W. T. 1986, {\it Numerical Recipes}, 
  (Cambridge:  Cambridge University Press)
\bibitem[Rubin \& Ford(1970)]{rub70} Rubin, V. C., \& Ford, W. K. 1970,
  \apj, 159, 379
\bibitem[Rubin \& Ford(1971)]{rub71} Rubin, V. C., \& Ford, W. K. 1971,
  \apj, 170, 25
\bibitem[Sato \& Sawa(1986)]{sat86} Sato, N. R., \& Sawa, T. 1986, \pasj, 
  38, 63
\bibitem[Sellwood(1985)]{sel85} Sellwood, J. A. 1985, \mnras,
217, 127
\bibitem[Stark(1977)]{sta77} Stark, A. A. 1977, \apj, 213, 368.

\bibitem[Tully \& Fisher(1977)]{tul77} Tully, R. B. \& Fisher, J. R.
1977, A\&A, 54, 661
\bibitem[Walterbos \& Kennicutt(1987)]{wal87} Walterbos, R. A. M., \&
  Kennicutt, R. C. 1987, \aaps, 69, 311 
\bibitem[Walterbos \& Kennicutt(1988)]{wal88} Walterbos, R. A. M., \&
Kennicutt, R. C. 1988, \aap, 198, 61 
\bibitem[Warren et al.(1992)]{war92} Warren, M. S., Quinn, P. J., 
 Salmon, J. K., \& Zurek, W. H. 1992, \apj, 399, 405
\bibitem[Widrow(2000)]{wid00} Widrow, L. M. 2000, \apjs, 131, 39
\bibitem[Zhao(1997)]{zha97} Zhao, H. 1997, 287, 525

\end{thebibliography}
\end{document}